\documentclass[aps,pra,groupedaddress,showpacs]{revtex4}
\pdfoutput=1
\usepackage{amsmath,amsfonts,amssymb}
\usepackage{epsfig}

\def\e{\begin{equation}}
\def\f{\end{equation}}
\def\^#1{{\bf #1}_0}
\def\=#1{\overline{\overline{#1}}}
\def\_#1{{\bf #1}}
\let\oslashed\o
\def\o{\omega}
\def\E{\varepsilon}
\def\M{\mu}
\def\.{\cdot}
\def\x{\times}

\def\l#1{\label{eq:#1}}
\def\r#1{(\ref{eq:#1})}
\def\d{\partial}
\def\D{\nabla}

\def\l#1{\label{eq:#1}}
\def\r#1{(\ref{eq:#1})}
\def\bra#1{\langle{#1}|}
\def\ket#1{|{#1}\rangle}

\newcommand\vect[1]{\left(\!\!\begin{array}{c}#1\end{array}\!\!\right)}
\newcommand\matr[1]{\left(\!\!\begin{array}{cc}#1\end{array}\!\!\right)}

\begin{document}

\title{Casimir repulsion in moving media}

\author{Stanislav I. Maslovski} \email{stas@co.it.pt}
\affiliation{
  Departamento de Engenharia Electrot\'{e}cnica\\
  Instituto de Telecomunica\c{c}\~{o}es, Universidade de Coimbra\\
  P\'{o}lo II, 3030-290 Coimbra, Portugal}

\date{\today}

\begin{abstract}
  Casimir-Lifshitz interaction emerging from relative movement of
  layers in stratified dielectric media (e.g., non-uniformly moving
  fluids) is considered. It is shown that such movement may result in
  a repulsive Casimir-Lifshitz force exerted on the layers, with the
  simplest possible structure consisting of three adjacent layers
  of the same dielectric medium, where the middle one is stationary
  and the other two are sliding along a direction parallel to the
  interfaces of the layers.
\end{abstract}


\maketitle
\section{Introduction}

In this paper we consider Casimir-Lifshitz forces
\cite{Casimir_attraction_PKNAW_1948, Lifshitz_force_SPJETP_1956} in
layered moving media. Our interest to this problem was initiated by a
recent discussion on the friction forces that may or may not appear
due to quantum-electromagnetic fluctuations in systems involving
moving dielectric
slabs~\cite{Leonhardt_nofriction_NJP_2009,Pendry_friction_NJP_2010,Leonhardt_comment_NJP_2010,
  Pendry_reply_NJP_2010,Barton_friction_NJP_2010,Hoye_friction_EPL_2010}.
In this paper, however, we will concentrate on another interesting
theoretical issue which, to the best of our knowledge, has not been
addressed so far: on the possibility of having {\it repulsive}
Casimir-Lifshitz forces in moving dielectrics. The so-called Casimir
repulsion is known to appear between electrically and magnetically
polarizable objects in vacuum
\cite{Boyer_vdWforces_PRA_1974,Santos_zeta_PRD_1999,Kenneth_repulsion_PRL_2002,
  Rosa_repulsion_PRD_2010}, or between dielectric objects of different
permittivity that are immersed in a dielectric fluid of an
intermediate permittivity
\cite{Dzyaloshinski_vdWforce_1961,Munday_repulsion_Nature_2009,Rahi_levitation_PRL_2010}. Very
recently, ultralong-range repulsive forces in piston configurations
involving cut metallic nanorods have been
reported~\cite{Maslovski_piston_PRA_2011}. There have been also
attempts on achieving repulsion or ``quantum levitation'' with the use
of other metamaterials~\cite{Henkel_casimir_EPL_2005,
  Leonhardt_levitation_NJP_2007,
  Pirozhenko_repulsion_JPA_2008,Rosa_metacasimir_PRL_2008,
  Rosa_repulsion_JPCS_2009,
  Yannopapas_repulsion_PRL_2009,Zhao_chiralrep_PRL_2010,
  Zhao_chiralrep_PRB_2010}. However, recently it has been shown that
the force between metal-dielectric metamaterial slabs in vacuum is
always attractive
\cite{Silveirinha_norepulsion_PRB_2010,Silveirinha_restrictions_PRA_2010,
  Silveirinha_chiralcomment_PRL_2010}. The symmetry considerations
also impose restrictions on the sign of the Casimir force
\cite{Kenneth_oppositesattract_PRL_2006,Rahi_equilibria_PRL_2010}.

In this paper we are going to consider the case in which the force
appears {\em only as the result of relative movement} of dielectric
layers. In contrast to the Casimir friction studies, we are interested
in the force component perpendicular to the direction of the
movement. The main idea of this work is to consider a system which is
initially balanced, i.e., when there is no movement there are no
fluctuation-induced forces. One example of a system with such property
is a uniform medium, say a fluid, which is initially at rest. There
is, however, a possibility that when separate layers of a fluid begin
to slide one with respect to another the balance is destroyed and
there appears a noncompensated attractive or repulsive interaction
between the sliding layers.  It should be well understood at this
point that the situation that we consider in this paper differs
principally from the previously studied case of moving dielectric
slabs {\it separated by a vacuum} \cite{Leonhardt_nofriction_NJP_2009,
  Pendry_friction_NJP_2010}. In the latter case, with an appropriate
Lorentz transformation for the electromagnetic field, one may always
reduce a problem involving a moving slab of an isotropic dielectric
{\em in vacuum} to an equivalent problem with a {\em stationary} slab
of the same isotropic dielectric in vacuum. This is possible because
under a Lorentz boost the vacuum ``background'' remains itself. Quite
differently, in this paper we study the Casimir-Lifshitz interactions
that appear in non-uniformly moving {\em matter}. Applying Lorentz
transformations in this case results in a more difficult problem
involving layers of anisotropic and nonreciprocal media.

Therefore, we are going to approach this problem without resorting to
an assumption that the available theories
\cite{Casimir_retardation_Pr_1948,Casimir_attraction_PKNAW_1948,
  Lifshitz_force_SPJETP_1956,VanKampen_VdWforces_PLA_1968} of the
Casimir-Lifshitz forces in dielectrics are also applicable in the case
of moving media. Instead, we quantize the electromagnetic field in
moving matter and derive a relation for the zero-point energy from the
first principles. This is required because moving media are not
invariant under time reversal and the traditional quantization scheme
based on a modal expansion in a large box is not applicable (at least,
without significant modifications). In fact, in this work we develop
an alternative quantization approach that allows to reuse many of the
results of the classic treatment of such nonreciprocal
media. Nevertheless, the results of our method fully agree with the
phenomenological quantization schemes developed by other authors
\cite{Jauch_quantization_PR_1948,Kong_quantization_JAP_1970,
Matloob_quantization_PRA_2005}.

The nonreciprocity considered in this paper is twofold: it may either
be a result of relativistic movements of material fluids or it may
manifest itself in uniaxial bianisotropic metamaterials the
constitutive relations of which include a term that is responsible for
nonreciprocal magnetoelectric coupling. Such metamaterials have been
theoretically known for a long time
\cite{Kamenetskii_swastika_ACEM_1997,Tretyakov_magnetoelectric_JEWA_1998,
  Tretyakov_fieldtransform_NJP_2008}; certain practical realizations
have been proposed as well
\cite{Tretyakov_moving_Metamaterials_2009}. Some authors do not make a
clear distinction between the real moving media and their metamaterial
counterparts, taking for granted that the two types can be described
with the constitutive relations of the same form. This is, however,
not entirely true. Although applying the Lorentz transformations to
the Maxwell equations written for a moving dielectric results (in the
laboratory frame) in bianisotropic material relations with
nonreciprocal magnetoelectric coupling, such a transformation may not
always lead to {\em spatially local} constitutive relations. Indeed,
the Lorentz transformation intermixes the spatial coordinates with
time, therefore, a medium which is nonlocal {\em in time} in one of
the reference frames (i.e., a dispersive dielectric in its proper
frame) becomes nonlocal {\em in both space and time} in another
reference frame. Thus, a moving dispersive dielectric may be described
(in the laboratory frame) with the equivalent spatially local
bianisotropic material relations only in a limited frequency range
where the dispersion is negligible.

Therefore, in this work the emphasis is mostly on weakly dispersive
moving magnetodielectrics for which one may assume that $\E(\o)$ and
$\M(\o)$ are practically constant and real in a wide range of
frequencies. This simplification, however, is not crucial for the main
theoretical prediction of this paper, namely, the existence of
repulsive Casimir-Lifshitz forces in layered moving media. This can be
seen from the known fact (see, e.g. \cite{Rosa_metacasimir_PRL_2008})
that the range of frequencies that make the dominant contribution to
the Casimir energy in a pair of material layers separated by the
distance $d$ is limited by $|\omega_{\rm max}| \approx 2\pi v_{\rm
  ph}/d$, where $v_{\rm ph}$ is the phase velocity in the background
material.  Thus, if $\o_{\rm max}$ is set to the upper boundary of the
region of low dispersion of a medium, then the theory developed in
this paper will apply at separations $d \gtrsim 2\pi v_{\rm
  ph}/|\o_{\rm max}|$. As there exist real materials with low
dispersion and loss up to, at least, the ultraviolet band, the
applicability range of our theory may start at hundreds of
nanometers. A straightforward generalization of the theory to the
dispersive case is outlined in one of the appendices.

The paper is organized as follows. In Section~\ref{waves} we solve
classically for the eigenwaves in a moving nondispersive medium and
discuss their properties. In Section \ref{hamiltonian} we derive an
expression for the Hamiltonian of the free electromagnetic field in a
moving medium and prove an orthogonality relation that holds for the
eigenmodes in such a nonreciprocal medium. In
Section~\ref{quantization} we quantize the (macroscopic)
electromagnetic fields in a moving medium and express the Hamiltonian
of the electromagnetic field in terms of the creation and annihilation
operators of a bosonic field. In Section~\ref{lifshitz} we obtain an
expression for the zero-point energy and its regular part that
represents the Casimir-Lifshitz interaction energy.  In
Section~\ref{layered} we solve for the Casimir-Lifshitz force in
layered moving media. In Section~\ref{numerics} we present and discuss
some numerical results that clearly demonstrate existence of repulsive
Casimir-Lifshitz forces in such media.

\section{\label{waves}Electromagnetic waves in a moving medium}

We consider a uniaxial medium (the axis is along $\^z$) which is
characterized by material relations of the following form [in this
section we work in the frequency domain; the time dependence is of the
form $\exp(-i\o t)$]:
\begin{align}
\l{matrel1}
\_D &= \=\E\.\_E + a\^z\x\_H,\\
\l{matrel2}
\_B &= \=\M\.\_H - a\^z\x\_E,
\end{align}
where $\=\E=\E_{\rm t}\=I_{\rm t}+\E\^z\^z$ and $\=\M=\M_{\rm
  t}\=I_{\rm t}+\M\^z\^z$ are the dyadic permittivity and the
permeability, respectively, with $\=I_{\rm t}$ being the unity dyadic
in the plane transversal to $\^z$, and $a$ is the parameter of
magnetoelectric coupling. Notice that due to the choice of signs in
\r{matrel1}--\r{matrel2} this coupling in nonreciprocal.

Such a medium can be envisioned either as a metamaterial with
nonreciprocal bianisotropic inclusions, or as a an effective medium
resulting from application of the Lorentz transformations to the
electromagnetic fields in a magnetodielectric moving with certain
velocity $v$ along the $z$-axis. In the latter case, the material
parameters as seen in the stationary frame satisfy (see,
e.g.,~\cite{Pauli_relativity_1958})
\begin{align}
\l{lor1}
\E_{\rm t} &= \E{1-\beta^2\over 1-n^2\beta^2},\\
\M_{\rm t} &= \M{1-\beta^2\over 1-n^2\beta^2},\\
\l{lor2}
a &= {\beta\over c}{n^2-1\over 1-n^2\beta^2},
\end{align}
where $\E$ and $\M$ are the permittivity and the permeability in the
comoving frame, $c=1/\sqrt{\E_0\M_0}$ is the speed of light in vacuum,
$\beta = v/c$, and $n^2=\E\M/(\E_0\M_0)$. The material parameters are
assumed nondispersive and lossless in \r{lor1}--\r{lor2}, but, in
fact, these relations may be also generalized for dispersive moving
media if plane waves are considered (this is further discussed in
Appendix~\ref{AppC}).

It should be noted that when these transformations are applied to a
medium with $n^2=1$, they result in $a = 0$ and the old values of the
permittivity and permeability, independently of the velocity
$v$. Thus, due to~\r{lor1}--\r{lor2}, a vacuum appears as a ``medium''
with properties invariant with respect to relative motion, while media
with nontrivial refractive index are seen differently in different
inertial frames of reference.

The Maxwell equations for the fields in a moving medium can be written as
\begin{align}
\l{max1}
i\o\=\M\.\_H &= \D_{\rm t}\x\_E + (i\o a + {\d_z})\^z\x\_E,\\
\l{max2}
-i\o\=\E\.\_E &= \D_{\rm t}\x\_H +(i\o a + {\d_z})\^z\x\_H,
\end{align}
where $\D_{\rm t} \equiv \=I_{\rm t}\.\D$ and $\d_z \equiv \d/\d
z$. Seeking for plane wave solutions of \r{max1}--\r{max2}, it is
possible to reduce Eqs.~\r{max1}--\r{max2} to
\begin{align}
\l{TMz}
  \left[\o^2\E_{\rm t}\M_{\rm t}-(k_z+\o a)^2-{\E_{\rm t}\over
      \E}k_{\rm t}^2\right]E_z &= 0, \quad(\mbox{TM}_z),\\
\l{TEz}
  \left[\o^2\E_{\rm t}\M_{\rm t}-(k_z+\o a)^2-{\M_{\rm t}\over
      \M}k_{\rm t}^2\right]H_z &= 0, \quad(\mbox{TE}_z),
\end{align}
where $\_k=\_k_{\rm t}+k_z\^z\^z$, $\_k_{\rm t}\equiv\=I_{\rm t}\.\_k$, is the wave
vector of a plane wave, and the two equations~\r{TMz} and~\r{TEz} are
for two independent polarizations: the transverse magnetic
polarization with respect to the $z$-axis (TM$_z$), for which
$H_z\equiv 0$, and the transverse electric polarization (TE$_z$), for
which $E_z\equiv 0$. The transversal components of the electric and
magnetic fields (with respect to the $z$-axis) in both TM$_z$ and
TE$_z$ polarizations can be expressed through the $z$-components of
the fields:
\begin{align}
\l{TMzH}
\_H_{\rm t}^{\rm TM_z} &= {\o\E(\_k_{\rm t}\x\^z)\over k_{\rm t}^2}E_z,\quad
\_E_{\rm t}^{\rm TM_z} = -{\E(k_z+\o a)\_k_{\rm t}\over\E_{\rm t}k_{\rm t}^2}E_z,\quad &(\mbox{TM}_z),\\
\l{TEzE}
\_E_{\rm t}^{\rm TE_z} &= -{\o\M(\_k_{\rm t}\x\^z)\over k_{\rm t}^2}H_z,\quad
\_H_{\rm t}^{\rm TE_z} = -{\M(k_z+\o a)\_k_{\rm t}\over\M_{\rm t}k_{\rm t}^2}H_z,\quad &(\mbox{TE}_z).
\end{align}

The electric displacement $\_D$ and the magnetic induction $\_B$ in
the same modes can be found with the help of the material
relations~\r{matrel1}--\r{matrel2} and the
relations~\r{TMzH}--\r{TEzE}:
\begin{align}
\begin{split}
\l{TMzBD}
\_B_{\rm t}^{\rm TM_z} &= {\E((\E_{\rm t}\M_{\rm t}-a^2)\o-ak_z)(\_k_{\rm t}\x\^z)\over \E_{\rm t}k_{\rm t}^2}E_z,\quad B_z^{\rm TM_z} = 0,\\
\_D_{\rm t}^{\rm TM_z} &= -{\E k_z\_k_{\rm t}\over k_{\rm t}^2}E_z, \quad D_z^{\rm TM_z} = \E E_z,
\end{split}
\quad &(\mbox{TM}_z),\\
\begin{split}
\l{TEzDB}
\_D_{\rm t}^{\rm TE_z} &= -{\M((\E_{\rm t}\M_{\rm t}-a^2)\o-ak_z)(\_k_{\rm t}\x\^z)\over \M_{\rm t}k_{\rm t}^2}H_z,\quad D_z^{\rm TE_z} = 0,\\
\_B_{\rm t}^{\rm TE_z} &= -{\M k_z\_k_{\rm t}\over k_{\rm t}^2}H_z, \quad B_z^{\rm TE_z} = \M H_z,
\end{split}
\quad &(\mbox{TE}_z).
\end{align}
An interesting property of the $\_D$ and $\_B$ vectors in a moving
medium is that despite the fact that the medium is anisotropic the
three vectors $\_k$, $\_D$, and $\_B$ are mutually orthogonal in each
of the TM$_z$ and TE$_z$ modes.

In the nondispersive case the dispersion equations \r{TMz}--\r{TEz}
are quadratic with respect to the frequency and can be easily
solved. As follows from \r{lor1}--\r{lor2} and \r{TMz}--\r{TEz} the
equations are the same for both TM$_z$ and TE$_z$ modes. The roots of
the dispersion equations are given by
\begin{align}
\l{roots}
{\o_{1,2}(\_k)\over c} = {\beta(n^2-1)k_z\pm
\sqrt{(1-\beta^2)[(n^2-\beta^2)k_{\rm t}^2+n^2(1-\beta^2)k_z^2]}
\over n^2-\beta^2}.
\end{align}
The expression under the square root is nonnegative because $n^2\ge 1$
and $\beta^2\le 1$. In the case when $\beta^2n^2 < 1$ (i.e., when the
velocity is below the threshold of the Cherenkov effect) only a single
solution of the dispersion equation is nonnegative, namely, the one with
the plus sign in~\r{roots}.

When $\beta^2n^2 \ge 1$ there may exist zero, one, or two nonnegative
roots of the dispersion equation, depending on the wave
vector. Without any loss of generality we may assume $v > 0$ and,
thus, $\beta > 0$. Then, the roots of the dispersion equation are both
negative (positive) if $k_z < 0$ ($k_z > 0$) and $|k_z/k_{\rm t}| >
{\sqrt{1-\beta^2\over\beta^2n^2-1}}$. When $|k_z/k_{\rm t}| <
{\sqrt{1-\beta^2\over\beta^2n^2-1}}$, there are two roots of opposite
signs. The boundary between these regions defines the Cherenkov cone
as seen from the stationary frame. In the comoving frame (i.e., in the
frame in which the medium is at rest), the same cone is seen as having
the half-angle $\theta$ such that $\tan\theta =
{1\over\sqrt{\beta^2n^2-1}}$, which is a well-known result.

The nonreciprocity of the material relations \r{matrel1}--\r{matrel2}
results in an obvious property of Eqs.~\r{TMz}--\r{TEz}: these
equations are not invariant with respect to the change of sign
of~$\o$. However, the time-harmonic fields that are the solutions of
\r{max1}--\r{max2} must satisfy the reality condition $\_F_{-\o}(\_x)
= \_F^*_{\o}(\_x)$, where $\_F$ represents either $\_E$ or $\_H$ and
the symbol~$^*$ denotes complex conjugation. Thus, their spatial
Fourier transforms, $\_F_{\o}(\_k) = \int
d^3\_x\,\_F_{\o}(\_x)e^{-i\_k\.\_x}$, that represent the complex
amplitudes of the respective plane waves, are such that
$\_F_{-\o}(-\_k) = \_F^*_{\o}(\_k)$. It is immediately seen that the
equations~\r{TMz}--\r{TEzDB} are invariant under such a transformation
that changes the signs of $\o$ and $\_k$ simultaneously. In addition
to this, the dispersion equations are also invariant with respect to a
simultaneous change of signs of $\o$ and $k_z$, as follows from~\r{TMz}--\r{TEz}.

The instantaneous fields in a given polarization, $\_F(\_x,t)$, can be
written as a superposition of the plane wave solutions of
\r{TMz} or \r{TEz}: \e \l{instF} \_F(\_x,t) = \int {d^3\_k\over
  (2\pi)^3}\sum_{p}\, \_F_{\o_p}(\_k)e^{i(\_k\.\_x-\o_p t)}, \f where
the index $p = 1,2$ labels the roots $\o_p=\o_p(\_k)$ [Eq.~\r{roots}]
for a given $\_k$, and $\_F_{\o_p}(\_k)$ represent the complex
amplitudes of the waves that belong to the two different branches
of~\r{roots}.

The reality condition $\_F_{-\o}(-\_k) = \_F^*_{\o}(\_k)$ allows to
rewrite~\r{instF} as follows.  We notice that the two branches
of~\r{roots} are such that $\o_1(\_k) = -\o_2(-\_k)$, and $\o_2(\_k) =
-\o_1(-\_k)$. Hence, by replacing $\_k$ with $-\_k$ in one of
the addends of the sum in~\r{instF}, Eq.~\r{instF} can be written in
the following equivalent form where only a single branch occurs
explicitly: \e \l{instF2} \_F(\_x,t) = \int {d^3\_k\over
  (2\pi)^3}\,\left[\_F_{\o}(\_k)e^{i(\_k\.\_x-\o t)}
  +\_F_{\o}^*(\_k)e^{-i(\_k\.\_x-\o t)}\right]. \f Any branch may be
chosen; for the following we select the branch with the plus sign in
front of the square root in \r{roots}.

\section{\label{hamiltonian}The Hamiltonian of the free electromagnetic field}

Classically, the Hamiltonian of the free electromagnetic field in a moving
medium can be obtained by considering the Maxwell equations written
for instantaneous fields:
\begin{align}
\l{maxt1}
{\d_t\_B} &= -\D\x\_E,\\
\l{maxt2}
{\d_t\_D} &= \D\x\_H,
\end{align}
where $\d_t\equiv \d/\d t$. Performing the standard steps on
derivation of the Poynting theorem, we write \e
\D\.(\_E\x\_H)=(\D\x\_E)\.\_H - (\D\x\_H)\.\_E = -(\d_t\_B)\.\_H -
(\d_t\_D)\.\_E.  \f Next, we use the material relations
\r{matrel1}--\r{matrel2} to express $\_D$ and $\_B$ in terms of $\_E$
and $\_H$ and, after recollecting the terms on the right-hand side
with some trivial vector algebra, we obtain \e \D\.(\_E\x\_H) =
-\d_t\left[{\_B\.\_H\over 2} + {\_D\.\_E\over 2}\right], \f i.e., the
same final result as in a stationary medium. Thus, the Hamiltonian is
(the same expression was used in~\cite{Kong_quantization_JAP_1970}) \e
\l{ham} {\cal H} = \int d^3\_x\left[{\_B\.\_H\over 2} + {\_D\.\_E\over
    2}\right] = \int{d^3\_k\over (2\pi)^3}
\left[{\_B(\_k,t)\.\_H(-\_k,t)\over 2} + {\_D(-\_k,t)\.\_E(\_k,t)\over
    2}\right],\f where $\_F(\_k,t)$ (with $\_F$ representing any of
the fields) are the time-dependent spatial Fourier transforms defined
by \r{instF}: \e \_F(\_k,t) = \sum_p \_F_{\o_p}(\_k)e^{-i\o_p t}. \f
Now we substitute the above representation into~\r{ham} and obtain

\begin{multline}
\l{hamk}
{\cal H} = {1\over 2}\int{d^3\_k\over (2\pi)^3} \sum_{p,s}
\left[\_B_{\o_p}(\_k)\.\_H_{\o_s}^*(\_k) + \_D_{\o_s}^*(\_k)\.\_E_{\o_p}(\_k)\right]e^{-i(\o_p-\o_s)t} = \\
={1\over 2}\int{d^3\_k\over (2\pi)^3}\sum_{p}
\left[\_B_{\o_p}(\_k)\.\_H_{\o_p}^*(\_k) + \_D_{\o_p}^*(\_k)\.\_E_{\o_p}(\_k)\right]+\\
+{1\over 2}\int{d^3\_k\over (2\pi)^3}
\sum_{p,s\atop p\neq s}
\left[\_B_{\o_p}(\_k)\.\_H_{\o_s}^*(\_k) + \_D_{\o_s}^*(\_k)\.\_E_{\o_p}(\_k)\right]e^{-i(\o_p-\o_s)t}.
\end{multline}
Since the dispersion equations are the same for both TE$_z$ and TM$_z$
modes, the field vectors that appear in \r{hamk} may be
regarded as arbitrary linear combinations of the fields of these two
main polarizations.

In an isolated conservative system the expression \r{hamk} represents
the total electromagnetic energy that remains constant when the system
evolves with time. Therefore, the last integral term in \r{hamk} that
explicitly depends on time must vanish. Using the Maxwell equations
written for plane waves, this term can be expressed as
\begin{multline}
\l{badterm}
{1\over 2}\int{d^3\_k\over (2\pi)^3}
\sum_{p,s\atop p\neq s}
\left[\_B_{\o_p}(\_k)\.\_H_{\o_s}^*(\_k) +
  \_D_{\o_s}^*(\_k)\.\_E_{\o_p}(\_k)\right]e^{-i(\o_p-\o_s)t}=\\
={1\over 2}\int{d^3\_k\over (2\pi)^3}
\sum_{p,s\atop p\neq s}
{\o_p+\o_s\over k^2}\,\_k\.[\_D_{\o_s}^*(\_k)\x\_B_{\o_p}(\_k)]e^{-i(\o_p-\o_s)t},
\end{multline}
where $k \equiv |\_k|$.
It is immediately seen that in a reciprocal medium this integral
vanishes for arbitrary Fourier transformed fields, because in such a
medium $\o_s(\_k) = \o_s(-\_k) = -\o_p(\_k)$.

In the moving medium, however, the situation is more
complicated. Consider, for example, the case when at $t = 0$ the
electromagnetic field forms a pulse composed of waves with the wave
vectors concentrated around $\_k=+\_k_0$ and $\_k=-\_k_0$. This
situation corresponds to defining an initial condition for the fields
in terms of an oscillating function (oscillating in space!) with a
smoothly varying amplitude vanishing at infinity. Then, in this pulse
there are waves with frequencies concentrated around
$\o=\pm\o_1(\_k_0)$ and $\o=\pm\o_2(\_k_0)$, whereas
$\o_1(\_k_0)\neq-\o_2(\_k_0)$. Let us look closer at the term
\r{badterm} in this case. We may get rid of the integration around
$\pm \_k_0$ in \r{badterm} because the spectral width of the pulse is
assumed to be small. Dropping an insignificant constant factor we
obtain
\begin{multline}
{1\over 2}\sum_{p,s\atop p\neq s}\sum_{\_k=\pm\_k_0}
{\o_p+\o_s\over k^2}\,\_k\.[\_D_{\o_s}^*(\_k)\x\_B_{\o_p}(\_k)]
e^{-i(\o_p-\o_s)t}=\\
={1\over 2}\sum_{p,s\atop p\neq s}{\o_p+\o_s\over k_0^2}
\_k_0\.[\_D_{\o_s}^*(\_k_0)\x\_B_{\o_p}(\_k_0) + \_D_{\o_p}(\_k_0)\x\_B_{\o_s}^*(\_k_0)]
e^{-i(\o_p-\o_s)t} =\\
={\o_1+\o_2\over k_0^2}
\,\mbox{Re}
\left\{
\_k_0\.[\_D_{\o_2}^*(\_k_0)\x\_B_{\o_1}(\_k_0) + \_D_{\o_1}(\_k_0)\x\_B_{\o_2}^*(\_k_0)]
e^{-i(\o_1-\o_2)t}
\right\}.
\end{multline}
The only possibility to make this term independent of time is to have
its amplitude vanishing: \e \l{orth}
\_k_0\.[\_D_{\o_2}^*(\_k_0)\x\_B_{\o_1}(\_k_0) +
\_D_{\o_1}(\_k_0)\x\_B_{\o_2}^*(\_k_0)] = 0.  \f It can be verified by
direct substitution that this condition holds for both TE$_z$ and
TM$_z$ modes (and also for any linear combination of them). Because
the same transformation that we have done above could be applied
directly to the integrand of \r{badterm}, we have proven that the term
\r{badterm} vanishes in general. Physically, Eq.~\r{orth} has the
meaning of an orthogonality condition for the modes with the wave
vector $\_k_0$ and the frequencies $\o_{1,2}(\_k_0)$ in a moving
medium.

Thus, we have proven that the Hamiltonian in a lossless non-dispersive
moving medium can be written as
\begin{multline}
  \l{hamfinal} {\cal H} = {1\over 2}\int{d^3\_k\over (2\pi)^3}\sum_{p}
  \left[\_B_{\o_p}(\_k)\.\_H_{\o_p}^*(\_k) + \_D_{\o_p}^*(\_k)\.\_E_{\o_p}(\_k)\right] =\\
  =\int{d^3\_k\over
    (2\pi)^3}\sum_{p}{\o_p\over k^2}\,\_k\.[\_D_{\o_p}^*(\_k)\x\_B_{\o_p}(\_k)] = \int{d^3\_k\over
    (2\pi)^3}{\o\over k^2}\,\_k\.[\_D_{\o}^*(\_k)\x\_B_{\o}(\_k)] +
  \mbox{c.c.},
\end{multline}
where in the last equality only a single branch occurs as in
\r{instF2}, and ``c.c.'' denotes the complex conjugate of the first
term. The only difference of \r{hamfinal} from the same expression for
a reciprocal magnetodielectric is in that $\o(-\_k)\neq \o(\_k)$.

We can split the total electric and magnetic fields in the last
expression into the components corresponding to the TM$_z$ and TE$_z$
polarizations. To do this we notice that the electric displacement
vector of the TM$_z$ (TE$_z$) mode and the magnetic induction vector
of the TE$_z$ (TM$_z$) mode are collinear. Thus, the cross terms in
the vector product $\_D_{\o}^*(\_k)\x\_B_{\o}(\_k)$ do not
contribute to the Hamiltonian \r{hamfinal}. Therefore, we can write
\begin{equation}
  \l{hamsep} {\cal H} =
  \int{d^3\_k\over (2\pi)^3}{\o\over k^2}\left(\_k\.[\_D_{\o}^*(\_k)\x\_B_{\o}(\_k)]^{\rm TM_z} + \_k\.[\_D_{\o}^*(\_k)\x\_B_{\o}(\_k)]^{\rm TE_z}\right)
  + \mbox{c.c.},\\
\end{equation}
where the brackets $[\ldots]^{\rm TM_z}$, $[\ldots]^{\rm TE_z}$ denote
the separate contributions of the respective modes.

The relations \r{hamfinal} and \r{hamsep} have a clear physical
meaning. Indeed, the term
$(\_k/k)\.[\_D_{\o}^*(\_k)\x\_B_{\o}(\_k)]$ corresponds to the
momentum of a plane wave in the moving medium. From the other hand,
the energy $w$ and the momentum $p$ of a plane wave are related by $w
= (\o/k)p$. Therefore, Eqs.~\r{hamfinal}--\r{hamsep}
may be understood as a summation over the energies of all possible
plane waves.

\section{\label{quantization}Quantization of the electromagnetic field in a moving medium}

The quantization of electromagnetic field in a moving medium is a
well-established subject (at least, in the nondispersive case) and can
be performed within different frameworks: (i)~with the covariant
Lagrangian formalism of Ref.~\cite{Jauch_quantization_PR_1948},
(ii)~with the Heisenberg formalism of
Ref.~\cite{Kong_quantization_JAP_1970}, (iii)~with the Green
tensor-based formalism of
Ref.~\cite{Matloob_quantization_PRA_2005}. All these approaches agree
and lead in effect to the so-called canonical quantization of the
electromagnetic field.

Perhaps, the most intuitive approach is the one based on Heisenberg
formalism. Under this approach one starts with a classical expression
for the Hamiltonian ${\cal H}$ in terms of the instantaneous fields as
in \r{ham}. The field variables as functions of the position and time
that appear in the Hamiltonian are promoted to Hermitian operators
that satisfy certain commutation relations. The commutation relations
must be such that the equations of motion in the Heisenberg formalism
(here and in what follows the square brackets $[\.,\.]$ denote the
commutator of two operators: $[A,B] = AB - BA$)
\begin{align}
\d_t\_B(\_x,t) &= (i\hbar)^{-1}[\_B(\_x,t),{\cal H}],\\
\d_t\_D(\_x,t) &= (i\hbar)^{-1}[\_D(\_x,t),{\cal H}],
\end{align}
result in a system of partial differential equations identical in form
with the classic Maxwell equations. Such an equivalence exists because
the classic electromagnetic theory can be thought of as a theory of
quantum states of light with a very large number of photons (which are
bosons) in each state. Although not required in the time-harmonic
regime, for an arbitrary time evolution the above (sourceless)
equations must be complemented by $\nabla\.\_D=\nabla\.\_B = 0$.

In~\cite{Kong_quantization_JAP_1970} it was shown that the required
equal-time commutation relations can be written in terms the Cartesian
components of $\_B$ and $\_D$ as \e \l{comDB}
[D_i(\_x,t),B_j(\_x',t)] = -i\hbar \E_{ijk}\d_k\delta(\_x-\_x'), \f
where $\E_{ijk}$ is the Levi-Civita tensor, $\d_k\equiv \d/\d x_k$,
and $\delta(\_x)$ is the three-dimensional Dirac delta function. Here
and in what follows we use Einstein's notation in which a summation
over repeating indices is assumed. It is also assumed that all
components of $\_D$ commute in between themselves, as do the
components of $\_B$.

From the commutation relation \r{comDB} it is seen that the
noncommuting components of the field operators $\_B$ and $\_D$ are
mutually orthogonal. Let us show that indeed such commutation
relations lead to the Maxwell equations~\r{maxt1}--\r{maxt2}. First,
we rewrite the Hamiltonian of the electromagnetic field \r{ham} in
terms of only $\_B$ and $\_D$: \e \l{hamBD} {\cal H} = \int
d^3\_x\left[{\_B\.\=\eta\.\_B\over 2} + {\_D\.\=\xi\.\_D\over 2} -
  {\chi \^z\.(\_B\x\_D-\_D\x\_B)\over 2}\right], \f where $\=\eta
=(\E_{\rm t}\M_{\rm t}-a^2)^{-1}\E_{\rm t}\=I_{\rm t}+\M^{-1}\^z\^z$,
$\=\xi =(\E_{\rm t}\M_{\rm t}-a^2)^{-1}\M_{\rm t}\=I_{\rm
  t}+\E^{-1}\^z\^z$, and $\chi = a(\E_{\rm t}\M_{\rm t}-a^2)^{-1}$ are
the parameters of the material relations~\r{matrel1}--\r{matrel2}
transformed to the form
\begin{align}
\_E &= \=\xi\.\_D -\chi\^z\x\_B,\\
\_H &= \=\eta\.\_B + \chi\^z\x\_D.
\end{align}
The symmetry of $\=\xi$ and $\=\eta$ and the form of the last addend
under the integral \r{hamBD} provide that the Hamiltonian is a
self-adjoint (Hermitian) operator: ${\cal H}^\dagger = {\cal H}$ (here and
in what follows the symbol $^\dagger$ denotes Hermitian conjugation).

Then, calculating, for example, the commutator of $\_B$ and ${\cal H}$
we find
\begin{multline}
  [B_i,{\cal H}] = {1\over 2}\int d^3\_x[B_i(\_x'),(\xi_{\alpha\beta}D_\alpha(\_x)D_\beta(\_x) -
  \chi \E_{z\alpha\beta}(B_\alpha(\_x)D_\beta(\_x) + D_\beta(\_x)B_\alpha(\_x))]=\\
  ={1\over 2}\int
  d^3\_x\left(\xi_{\alpha\beta}([B_i(\_x'),D_\alpha(\_x)]D_\beta +
    D_{\alpha}[B_i(\_x'),D_\beta(\_x)]) -
    2\chi \E_{z\alpha\beta}B_\alpha(\_x)[B_i(\_x'),D_\beta(\_x)]\right)=\\
  ={i\hbar\over 2}\int
  d^3\_x\,\d_k\delta(\_x-\_x')\left[\xi_{\alpha\beta}(\E_{\alpha
      ik}D_\beta(\_x) + \E_{\beta ik}D_{\alpha}(\_x))
    - 2\chi\E_{z\alpha\beta}\E_{\beta ik}B_\alpha(\_x)\right]=\\
  =-i\hbar\int
  d^3\_x\,\delta(\_x-\_x')\E_{ik\alpha}\d_k\left[\xi_{\alpha\beta}D_\beta(\_x)-\chi\E_{\alpha
      z\beta}B_{\beta}(\_x)\right]= -i\hbar\E_{ik\alpha}\d_kE_\alpha,
\end{multline}
which is the same as $[\_B,{\cal H}] = -i\hbar\D\x\_E$. In the
derivation we used the fact that
$\xi_{\alpha\beta}=\xi_{\beta\alpha}$. In a similar manner one obtains
$[\_D,{\cal H}] = i\hbar\D\x\_H$.

The standard way to proceed after this step is to make a transition
into the momentum space by expressing $\_D(\_x)$ and $\_B(\_x)$ in
terms of a pair of conjugate canonical variables $\_P(\_k)$ and
$\_Q(\_k)$. One then diagonalizes the Hamiltonian written in terms of
$\_P(\_k)$ and $\_Q(\_k)$ by introducing the creation and annihilation
operators. We would like, however, to move along another way that will
allow us to reuse many of the results of the classic theory considered
in the previous sections.

To make a connection with the frequency domain treatment of Section
\ref{waves} we look for the solutions of the Maxwell equations
(written for the quantum vector field operators!) that have the form
(here $\_F$ represents any field vector)
\begin{equation}
\l{Fproj}
  \_F(\_x,t) = \int {d^3\_k\over (2\pi)^3}\,\_F_{\o}(\_k)e^{i[\_k\.\_x-\o(\_k) t]},
\end{equation}
where the operators $\_F_{\o}(\_k)$ can be understood as the (time and
position-independent) wave amplitude operators. The reality condition
requires $\_F(\_x,t)$ to be an Hermitian operator, thus
$\_F_{-\o}(-\_k) =\_F_{\o}^\dagger(\_k)$.  When such a form is
substituted into the Maxwell equations, one can reduce these equations
to \r{TMz}--\r{TEzDB} with all the field variables promoted to wave
amplitude operators.

As the wave amplitude operators are assumed non-trivial, the frequency
$\o(\_k)$ in \r{Fproj} is found by solving a dispersion equation which
is identical to the classic one. Thus, there are two dispersion
branches $\o_p(\_k) = -\o_s(-\_k)$, $s\neq p$, and, analogously to
\r{instF} and \r{instF2}, we can write \e \l{Fpxt}\_F(\_x,t) = \int
{d^3\_k\over (2\pi)^3}\,\left[ \_F_{\o}(\_k)e^{i(\_k\.\_x-\o t)} +
  \_F_{\o}^\dagger(\_k)e^{-i(\_k\.\_x-\o t)}\right], \f where only a single
branch $\o(\_k)$ appears explicitly [as before, we select the branch
with the positive square root in \r{roots}].


With enough care, the results of Section \ref{hamiltonian} may be also
promoted to operators, provided that they are written in a form that
satisfies the reality condition for the Hamiltonian: ${\cal H}^\dagger =
{\cal H}$. Thus, in the operator form the Hamiltonian~\r{hamfinal}
becomes
\begin{multline}
\l{hamfinop} {\cal H} =
{1\over 2}\int{d^3\_k\over
  (2\pi)^3}{\o\over k^2}\,\_k\.[\_D_{\o}^\dagger(\_k)\x\_B_{\o}(\_k) -
  \_B_{\o}(\_k)\x\_D_{\o}^\dagger(\_k)]  + \mbox{h.c.} =\\
={1\over 2}\int{d^3\_k\over
  (2\pi)^3}{\o\over k^2}\left[[\_k\x\_D_{\o}^\dagger(\_k)]\.\_B_{\o}(\_k) +
\_B_{\o}(\_k)\.[\_k\x\_D_{\o}^\dagger(\_k)]\right] + \mbox{h.c.},
\end{multline}
where ``h.c.'' stands for the Hermitian conjugate of the first term.

The representation of the Hamiltonian in terms of $\_D$ and $\_B$ is
useful because in both TM$_z$ and TE$_z$ modes in a moving medium the
three vectors $\_k$, $\_D$, and $\_B$ are mutually orthogonal, as has
been found in Section \ref{waves}. Therefore, in each mode separately
the vectors $\_B$ and $\_k\x\_D$ are collinear, while the same vectors
corresponding to the two different modes are mutually
orthogonal. Thus, with the help of Eqs.~\r{TMzBD}--\r{TEzDB} we may
express the vectors $\_B_{\o}(\_k)$ and $\_k\x\_D_{\o}(\_k)$ as
\begin{align}
\l{BinA}
  \_B_{\o}(\_k) &= \sqrt{ck\hbar\M_{\rm t}\over 2}\left(\gamma_0 a_1(\_k)\,\_e_1 + {a_2(\_k)\over \gamma_0 c\sqrt{\E_{\rm t}\M_{\rm t}}}\,\_e_2\right),\\
\l{kDinA}
  \_k\x\_D_{\o}(\_k) &= k\sqrt{k\hbar\over 2c\M_{\rm t}}\left({a_1(\_k)\over \gamma_0}\,\_e_1 + {\gamma_0 c\sqrt{\E_{\rm t}\M_{\rm t}}}a_2(\_k)\,\_e_2\right),
\end{align}
where $\gamma_0^2 = [(\E_{\rm t}\M_{\rm t}-a^2)\o-ak_z]/(ck\E_{\rm
  t}\M_{\rm t})$ (as everywhere above, we use the branch of \r{roots}
with the plus sign, therefore $\gamma_0^2 =
\sqrt{n^2(1-\beta^2)k_z^2+(n^2-\beta^2)k_{\rm
    t}^2}/(nk\sqrt{1-\beta^2}) \ge 0$), and $a_{1,2}(\_k)$ are the
amplitude operators of the TM$_z$ and TE$_z$ modes, respectively (the
coefficients in front of \r{BinA}--\r{kDinA} are to ensure that these
operators have the dimension of $\sqrt{\rm m^3}$). The unit vectors
$\_e_{1,2}$ are defined as $\_e_1=(\_k\x\_z_0)/|\_k\x\_z_0|$ and
$\_e_2=(\_k\x\_e_1)/|\_k\x\_e_1|$. The Hamiltonian~\r{hamfinop} can be
now expressed as
\begin{equation}
\l{hama}
  {\cal H} = {\hbar\over 2}\int{d^3\_k\over
    (2\pi)^3}\sum_{q}\o(\_k)\left[a_q^\dagger(\_k)a_q(\_k)+a_q(\_k)a_q^\dagger(\_k)\right],
\end{equation}
where the index $q=1,2$ labels the two main polarizations.

The operators $a_{1,2}(\_k)$ that we have introduced
above for the two main polarizations must satisfy certain commutation
relations that should in the end lead to the commutation relation
\r{comDB} for the quantum fields $\_D(\_x,t)$ and $\_B(\_x,t)$. We may write
\begin{align}
\begin{split}
\_B(\_x,t) = \sqrt{c\hbar\M_{\rm t}\over 2}\int{d^3\_k\,\gamma_0\sqrt{k}\over (2\pi)^3}\,\_e_1\left[
a_1(\_k)e^{i(\_k\.\_x - \o t)} + a_1^\dagger(\_k)e^{-i(\_k\.\_x - \o t)}
\right]+\\
+\sqrt{\hbar\over 2c\E_{\rm t}}\int{d^3\_k\,\sqrt{k}\over(2\pi)^3\gamma_0}\,\_e_2\left[
a_2(\_k)e^{i(\_k\.\_x - \o t)} + a_2^\dagger(\_k)e^{-i(\_k\.\_x - \o t)}
\right],
\end{split}\\
\begin{split}
\_D(\_x,t) = \sqrt{\hbar\over 2c\M_{\rm t}}\int{d^3\_k\over(2\pi)^3\gamma_0\sqrt{k}}\,(\_e_1\x\_k)\left[
a_1(\_k)e^{i(\_k\.\_x - \o t)} + a_1^\dagger(\_k)e^{-i(\_k\.\_x - \o t)}
\right]+\\
+\sqrt{c\hbar\E_{\rm t}\over 2}\int{d^3\_k\,\gamma_0\over(2\pi)^3\sqrt{k}}\,(\_e_2\x\_k)\left[
a_2(\_k)e^{i(\_k\.\_x - \o t)} + a_2^\dagger(\_k)e^{-i(\_k\.\_x - \o t)}
\right].
\end{split}
\end{align}
It can be verified that the commutation relation \r{comDB} follows
from these formulas if the operators $a_{1,2}(\_k)$ satisfy the
canonical commutation relations for annihilation and creation
operators for bosons:
\begin{align}
\l{comA0}
[a_i(\_k),a_j(\_k')] = [a_i^\dagger(\_k),a_j^\dagger(\_k')] &= 0,\\
\l{comA}
[a_i(\_k),a_j^\dagger(\_k')] = - [a_i^\dagger(\_k),a_j(\_k')] &= (2\pi)^3\delta_{ij}\delta(\_k-\_k'),
\end{align}
where $\delta_{ij}$ is Kronecker's delta.
Indeed, with the help of the above formulas one may write 
\begin{multline}
[D_i(\_x,t), B_j(\_x',t)] =
-{\hbar\over 2}\int{d^3\_k\over (2\pi)^3}\int{d^3\_k'\over (2\pi)^3}
{\sqrt{k'}\over \sqrt{k}}\bigg\{{\gamma_0'(\_k\x\_e_1)_i(\_e_1)_j\over\gamma_0}\x\\
\x\left(
[a_1(\_k),a_1^\dagger(\_k')]e^{i(\_k\.\_x - \_k'\.\_x' - \o t + \o't)} +
[a_1^\dagger(\_k),a_1(\_k')]e^{i(-\_k\.\_x + \_k'\.\_x' + \o t - \o't)}
\right)+\\
+{\gamma_0(\_k\x\_e_2)_i(\_e_2)_j\over\gamma_0'}\left(
[a_2(\_k),a_2^\dagger(\_k')]e^{i(\_k\.\_x - \_k'\.\_x' - \o t + \o't)} +
[a_2^\dagger(\_k),a_2(\_k')]e^{i(-\_k\.\_x + \_k'\.\_x' + \o t - \o't)}
\right)\bigg\},
\l{hugeexp}
\end{multline}
where $\gamma_0'\equiv \gamma_0(\_k')$ and
$\o'\equiv\o(\_k')$. Substituting \r{comA} into \r{hugeexp} and taking the integral
over $\_k'$ one obtains
\begin{multline}
[D_i(\_x,t), B_j(\_x',t)] = -{\hbar\over 2}\int{d^3\_k\over (2\pi)^3}
\big[\_k\x(\_e_1\_e_1+\_e_2\_e_2)\big]_{ij}\left(e^{i\_k\.(\_x-\_x')}-e^{-i\_k\.(\_x-\_x')}\right)=\\
= i\hbar\E_{i\alpha\beta}\delta_{\beta j}{\d\over \d x_\alpha}\int{d^3\_k\over (2\pi)^3}
e^{i\_k\.(\_x-\_x')} =-i\hbar\E_{ijk}\d_k\delta(\_x-\_x').
\end{multline}
In this derivation we used the fact that the vectors $\_e_{1,2}$ and
$\_k$ form a triplet of mutually orthogonal vectors, and, thus,
$\_k\x(\_e_1\_e_1+\_e_2\_e_2) = \_k\x\=I$, where $\=I$ is the unity
dyadic: $\big(\=I\big)_{ij} = \delta_{ij}$.

In a similar and simpler manner one can also check that the relations
\r{comA0}--\r{comA} ensure that all components of $\_B(\_x,t)$, as
well as all components of $\_D(\_x,t)$, commute among themselves.

Therefore,
following~\cite{Jauch_quantization_PR_1948,Kong_quantization_JAP_1970}
we may conclude that the quantization of the electromagnetic field in
a moving lossless nondispersive medium leads to the canonical result,
with all the field operators and the Hamiltonian expressed in terms of
the standard annihilation and creation operators of a bosonic
field.

\section{\label{lifshitz}The expression for the zero-point energy}
The canonical diagonalized form of the Hamiltonian \r{hama} allows for
introduction of the particle number operator ${\cal N}_{\_k,q}$. By
definition, the action of the number operator on a state results in
the number of photons in this state: $n_{\_k,q} =
\bra{\Psi_{\_k,q}}{\cal N}_{\_k,q}\ket{\Psi_{\_k,q}}$. However, in
order for this to hold the states must be properly defined and
normalized, so that $\langle\Psi_{\_k,q}|\Psi_{\_k,q}\rangle=1$. One
way to achieve this is to discretize the $\_k$-vector space into cells
of infinitesimal volumes $V_{\_k}$ centered around the points $\_k$
and require that in each state each cell contains an integral number
of photons.

Then, the number operator can be introduced as $ {\cal N}_{\_k,q}=
\int_{V_{\_k}}\!{d^3\_k'\over
  (2\pi)^3}a^\dagger_q(\_k')a_q(\_k')=\int_{V_{\_k}}\!{d^3\_k'\over
  (2\pi)^3}a^\dagger_q(\_k')a_q(\_k)=\int_{V_{\_k}}\!{d^3\_k'\over
  (2\pi)^3}a^\dagger_q(\_k)a_q(\_k')$. The last two equalities are
equivalent and hold because $V_{\_k}$ is infinitesimal. Then, from the
commutation relation \r{comA} it follows that the Hamiltonian \r{hama}
may be expressed in terms of the number operator as \e \l{hamN} {\cal
  H} = \sum_{\_k}\sum_{q}\hbar\o_{\_k,q}\left[{\cal N}_{\_k,q}+{1\over
    2}\right],\f where the first sum is taken over all the cells in
the wave vector space. In \r{hamN} we have labeled the frequency with
an index $q$ just to remind that the two main polarizations could in
principle have different dispersion (which is the case of Kong's
paper~\cite{Kong_quantization_JAP_1970} where the medium {\em at rest}
is assumed uniaxial).  The term \e \l{zpen}{\cal E} =
\sum_{\_k}\sum_{q}{\hbar\o_{\_k,q}\over 2} \f corresponds to the
so-called zero-point energy, i.e., to the energy of the ground state
of a quantum field.  The sum \r{zpen} is wildly divergent and must be
treated with a suitable renormalization procedure. It is known,
however, that besides being divergent the zero-point energy ${\cal E}$
may in some situations lead to physically observable phenomena, for
instance, it plays a key role in the physics of the Casimir-Lifshitz
forces.

In a typical scenario in which one may observe a force due to the
zero-point fluctuations of a quantum field, there exists a geometrical
parameter, $d$, that affects the modal dispersion and the density of
quantum states of a system. Hence, this parameter, by virtue of
\r{zpen}, also affects the zero-point energy: ${\cal E} = {\cal
  E}(d)$. Any slow rate, quasi-stationary variations in this parameter
result in variations in the amount of energy associated with the
quantum fluctuations, which means that there appears a macroscopic
force proportional to $\partial{\cal E}(d)/\partial d$.

For example, let us consider a layer of moving medium sandwiched in
between two perfectly electrically conducting (PEC) mirrors positioned
at $x = \pm d/2$. As before, we assume that the medium moves along the
$z$-axis, so that the introduced mirrors do not interfere with the
movement. It is evident that in this problem the modal spectrum is
discrete in $k_x$ (to see this one has to complement the field
equations \r{max1}--\r{max2} with the boundary conditions at the
mirrors), while $k_y$ and $k_z$ form a continuous spectrum. Therefore,
\r{zpen} may be written as \e \l{pecec}{{\cal E}\over L^2} =
\sum_{q}{\sum_{n}}'\int{dk_y\,dk_z\over (2\pi)^2}{\hbar\o_{({\pi
      n/d},k_y,k_z),q}\over 2}, \f where ${\cal E}/L^2$ has the
meaning of the energy in the considered cavity per unit area of the
mirrors; the infinite summation over $n\in\mathbb{Z}$ skips $n=0$ for
the TM$_z$ modes.

The frequencies that appear in the summation \r{pecec} may be
understood as the eigenfrequencies of a resonator formed by a layer of
a moving medium and the mirrors. In general, for a pair of polarization
sensitive, i.e., anisotropic mirrors (we will need this for the next
section) the modes of such a resonator can be found by introducing
$2\x2$ reflection matrices (or, in other terms, planar dyadics)
$\=R_{1,2}(\o,k_y,k_z)$ of the mirrors and the complex propagation
factor $\gamma(\o,k_y,k_z) \equiv -ik_x(\o,k_y,k_z)$ of the waves that
travel in between the mirrors. Then, in terms of these quantities the
characteristic equation for the modes in the cavity is readily
obtained as
\begin{equation}
\l{chareq}
  {\cal D}(\o,k_y,k_z,d)\equiv\det\left[\={I^{(2)}} - \=R_1(\o,k_y,k_z)\.\=R_2(\o,k_y,k_z)
    e^{-2\gamma(\o,k_y,k_z)d}\right] = 0,
\end{equation}
where $\={I^{(2)}}$ is the planar unity dyadic. The characteristic
equation \r{chareq} for the case of the ideally conducting mirrors
reduces to $(1-e^{-2\gamma d})^2 = 0$ with the obvious solution $k_x = \pi
n/d$ that appears in \r{pecec}.

In what follows we are going to use the principle of argument to
replace the discrete summation over the resonant frequencies in
\r{pecec} by an integration in the complex plane of~$\omega$. Indeed,
if a function~$f(\o)$ is analytic and has roots in a closed region $G$
with the boundary~$\partial G$, then the sum over its roots in this
region can be found as $\sum \o_k = (2\pi i)^{-1}\oint_{\partial G}\o
d\log f(\o)$. There is, however, a subtle difficulty when applying
this principle to the function of the characteristic equation
\r{chareq}, because ${\cal D}(\omega,k_y,k_z,d)$ may have poles and
branch points. The poles may appear at the points where the reflection
coefficients $\=R_{1,2}(\omega,k_y,k_z)$ have resonances:
$|\=R_{1,2}|\rightarrow \infty$, and, thus, they correspond to surface
waves that may exist at the boundaries between two different
media. Respectively, the branch points appear at the frequencies where
$\gamma(\omega,k_y,k_z) = 0$, i.e., at the points where the
propagating waves transition into the evanescent ones. The main
difficulty is with the branch points, as the poles of the
reflection coefficients do not depend on $d$ and only add a constant
to the sum representing the zero-point energy, i.e., they only
shift the origin of the zero point energy which is irrelevant.

However, it is possible to rewrite the characteristic equation in a
form that does not have branch points and is meromorphic in the
complex plane of $\o$ (see Appendix~\ref{AppB}). When such a form of the
characteristic equation is used (here we use the same symbol ${\cal
  D}$ to denote the function of this characteristic equation), the
summation over the discrete frequencies for a given pair of $k_y$,
$k_z$ in~\r{pecec} can be formally replaced with an integration over a
path $C$ in the complex plane of $\o$ that encircles the roots of the
characteristic equation:
\begin{equation}
\l{argpr}
  \sum_{q}{\sum_{n}}'{\hbar\o_{({\pi n/d},k_y,k_z),q}\over 2}=
  {\hbar\over 4\pi i}\int\limits_C\o\,d\log{\cal D}(\o,k_y,k_z,d).
\end{equation}
When $\beta^2n^2 < 1$ the roots we are interested in lie on the
positive half of the real axis (see Section~\ref{waves}), therefore we
can choose the path $C$ so that it follows the imaginary axis from
$+i\infty$ to $-i\infty$ and then closes in the right half of the
$\o$-plane with a semicircle $C_{\infty}$ of an infinite radius.

It should be well understood at this point that the integral \r{argpr}
diverges, as well as the original series \r{pecec} does. Nevertheless,
one may find a way to regularize \r{argpr} by dropping
distant-independent infinite terms in the integration \r{argpr}, as
explained in Appendix~\ref{AppB}. Doing this, the regular part of the
zero-point energy, or, in other terms, the Casimir interaction energy
at zero temperature, $\delta{\cal E}$, can be expressed with an
integral over the imaginary axis only:
\begin{multline}
\l{lifshitz}
{\delta{\cal E}\over L^2}= -{\hbar\over 4\pi i}\int{dk_y\,dk_z\over (2\pi)^2}
\int\limits_{-i\infty}^{+i\infty}\o\,d\log{\cal D}(\o,k_y,k_z,d)=
{\hbar\over 4\pi}\int{dk_y\,dk_z\over (2\pi)^2}\x\\
\x\int\limits_{-\infty}^{+\infty}\log{\cal D}(i\xi,k_y,k_z,d)\,d\xi =
{\hbar\over 2\pi}\int{dk_y\,dk_z\over (2\pi)^2}\int\limits_0^{+\infty}
\log{\cal D}(i\xi,k_y,k_z,d)\,d\xi,
\end{multline}
where we replaced the integration variable by $\o=i\xi$ and integrated
by parts once. The last equality holds due to the symmetry with
respect to a simultaneous change of signs of $\xi$ and $k_z$: ${\cal
  D}(-i\xi,k_y,-k_z) = {\cal D}(i\xi,k_y,k_z)$. One may recognize in
\r{lifshitz} the so-called generalized Lifshitz formula for the
Casimir energy at zero temperature.

As we are not using a covariant formulation of electrodynamics in this
paper, the relativistic covariance of the obtained result~\r{lifshitz}
requires an additional discussion. Some implications of special
relativity on the reflection matrices $\=R_{1,2}$ are outlined in
Appendix~\ref{AppA}. In particular, it can be verified that if there
exists a reference frame in which the moving media are at rest,
then~\r{lifshitz} reduces to the classic Dzyaloshinski-Lifshitz result
\cite{Dzyaloshinski_vdWforce_1961} for the Casimir energy of
stationary magnetodielectric slabs. This is, of course, just a
consequence of the material relation
transformations~\r{lor1}--\r{lor2}.

Above the threshold of the Cherenkov radiation, i.e., when $\beta^2n^2
> 1$, the dispersion relation~\r{roots} may result in negative
frequencies irrespectively of which branch of \r{roots} is
selected. By virtue of \r{hamN} this leads to the appearance of
negative quanta in the range of wave vectors that belong to the
Cherenkov cone. These quanta are responsible for a potential
instability in a medium that moves with a velocity higher than the
phase velocity in the same medium at rest. Indeed, any stationary (in
the laboratory frame) object that perturbs the electromagnetic field
will radiate in such quickly moving medium. Because of this unavoidable
instability, in the following sections we restrict our analysis only
by the case when $\beta^2n^2 < 1$.

\section{\label{layered}Casimir energy and force in layered moving media}

\begin{figure}
\centering
\epsfig{file=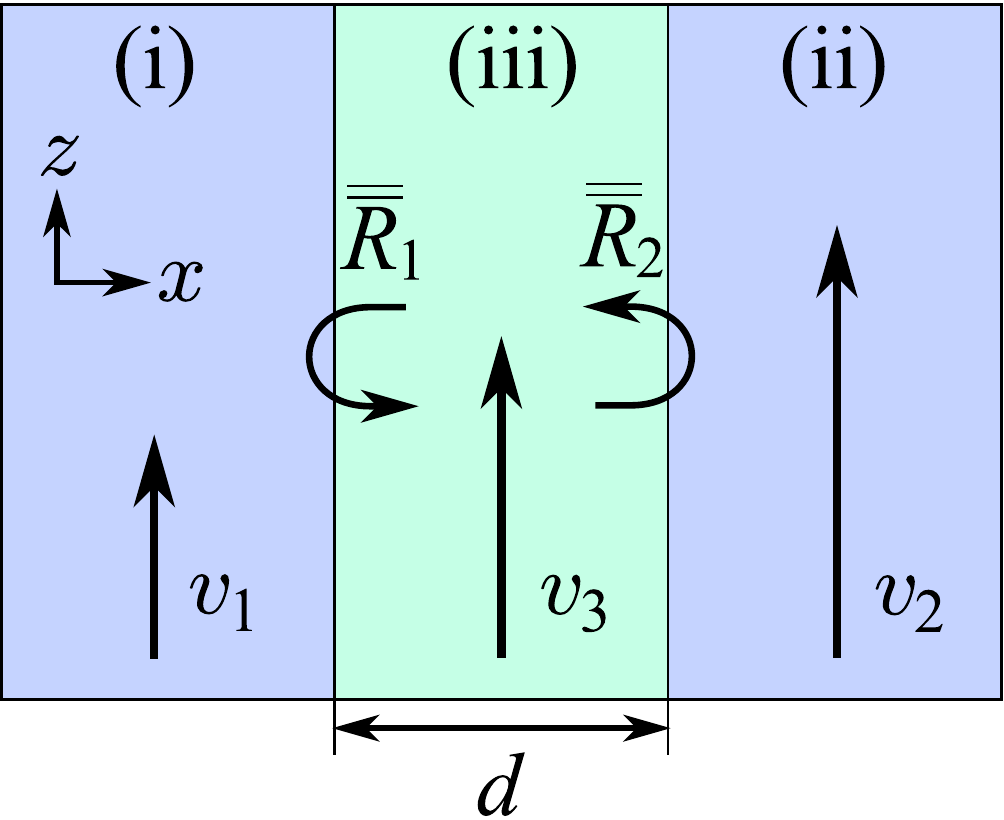,width=0.35\textwidth}
\caption{\label{fig_layers} (Color online) A layer of moving magnetodielectric (iii) of
  thickness $d$ sandwiched in between two moving semiinfinite
  magnetodielectric layers (i) and (ii). The three layers slide along
  the same line which is parallel to the interfaces of the layers. The
  magnitudes and the signs of the velocities $v_{1,2,3}$ are
  arbitrary. The Casimir-Lifshitz force is calculated from the
  reflection matrices $\overline{\overline{R}}_{1,2}$ defined at the
  interfaces of the layers.}
\end{figure}

In this section we will consider the Casimir energy and force that
result from the zero-point fluctuations in a layered moving medium,
namely, in a configuration analogous to the canonical problem solved
by
Lifshitz~\cite{Lifshitz_force_SPJETP_1956,Dzyaloshinski_vdWforce_1961}. Thus,
we consider a structure composed of a moving layer of finite thickness
$d$ sandwiched in between two other semiinfinite moving layers
(Fig.~\ref{fig_layers}). The velocities are assumed
uniform within the layers and collinear with the $z$-axis which is
parallel to the layer interfaces. Such a structure may be understood
as a simplified model of a nonuniformly moving fluid, in which the
width of the transition regions where the velocity changes
continuously is assumed small compared to the thickness of the
layers. In other words, we neglect all friction effects that may exist
at the boundaries of the moving layers. In practice such situation may
be achievable, for example, in certain phases of liquid helium at very
low temperatures or within metamaterial layers where the velocity is
merely a structural parameter (i.e., when there is no real movement).

It is evident that one may always choose a reference frame in which
the middle layer is at rest. However, we prefer not to impose such a
restriction, ensuring in this way a straightforward generalization of
our results to the case of multiple moving layers.
As follows from the treatment of Section~\ref{lifshitz}, in order to
obtain the Casimir energy of this system, one must first solve for the
reflection matrix at an interface of two moving layers. Although there
are some results available in the literature (see, e.g.,
\cite{Huang_movingrefl_JAP_1994} and references therein), they are
typically given in a form unsuitable for our purposes (e.g., one of
the layers is assumed to be vacuum) therefore, in Appendix~\ref{AppA} we derive
the necessary expressions for the components of the reflection
matrices
\begin{equation}
\l{reflmatrSec}
\=R_{1,2}\equiv \matr{
  R_{1,2}^{\rm ee} & R_{1,2}^{\rm eh}\\
  R_{1,2}^{\rm he} & R_{1,2}^{\rm hh} }
\end{equation}
that are defined in terms of the $z$-components of the electric and magnetic
fields. With these expressions at hand, the Casimir interaction energy in the
canonical triple-layer structure is given by \r{lifshitz} where the
matrices $\=R_{1,2}$ correspond to the two interfaces of the middle
layer. From the expressions derived in Appendix~\ref{AppA}, it is also seen
that the reflection matrices are invariant under a simultaneous change
of signs of $\o$ and $k_z$: we used this property when obtaining the
expression for the zero point energy~\r{lifshitz}.

Next, the Casimir force component normal to the
interfaces is found by differentiating~\r{lifshitz} with respect to
the thickness of the middle layer~$d$ (we use the convention that a
positive force corresponds to attraction):
\begin{multline}
\l{force1}
{F_{\rm c}\over L^2} = {\hbar\over 2\pi}\int{dk_y\,dk_z\over(2\pi)^2}
\int\limits_0^\infty{\d\over \d d}\log\det\left\{\={I^{(2)}}-\=R_1\.\=R_2\,e^{-2\gamma d}\right\}d\xi = \\
={\hbar\over 2\pi}\sum_{n=1}^2\int{dk_y\,dk_z\over(2\pi)^2}\int\limits_0^\infty
{2\lambda_n\gamma e^{-2\gamma d}\over 1 - \lambda_n e^{-2\gamma d}}\,d\xi,
\end{multline}
where $\lambda_{1,2}$ are the eigenvalues of the matrix
$\=R_1\.\=R_2$. It can be shown that the same expression for the Casimir force must
also hold in the case of dispersive material parameters which is
discussed in Appendix~\ref{AppC}.

Because the dispersion equation for the waves in (lossless and
nondispersive) moving media is not symmetric with respect to the
change of sign of the frequency $\o$, it is not anymore a function of
$\o^2$ as in (lossless and nondispersive) reciprocal media. Due to
this asymmetry the reflection matrices $\=R_{1,2}$ are in general {\em
  complex} at the imaginary frequencies $\o = i\xi$ while the
respective matrices in reciprocal media are always real under the same
circumstances. Therefore, in general, the eigenvalues $\lambda_{1,2}$
of the matrix $\=R_1\.\=R_2$ are also complex. In
Section~\ref{numerics} we will, however, show that the expression for
the Casimir force always results in real numbers, due to the symmetry
of the integrand of ~\r{force1}.

To simplify the integral \r{force1} further we introduce new dimensionless variables
$\kappa_y = ck_y/\xi$, $\kappa_z = ck_z/\xi$,
$\nu = c\gamma/\xi$, and  $\zeta = \xi d/c$, in which \r{force1} becomes
\begin{equation}
\l{force11}
{F_{\rm c}\over L^2}=
{\hbar c\over 2\pi d^4}\sum_{n=1}^2\int{d\kappa_y\,d\kappa_z\over(2\pi)^2}\int\limits_0^\infty
{2\nu\lambda_n\zeta^3 e^{-2\nu\zeta}\over 1 - \lambda_n e^{-2\nu\zeta}}\,d\zeta.
\end{equation}
One may notice that both $\lambda_n$ and $\nu$ do not depend on
$\zeta$ (they depend only on the relative wavenumbers $\kappa_{y}$ and $\kappa_z$ because the
material parameters are assumed nondispersive), therefore, by substituting $\zeta = t/(2\nu)$ we obtain
\begin{equation}
\l{force2}
{F_{\rm c}\over L^2}=
{\hbar c\over 16\pi d^4}\sum_{n=1}^2\int{d\kappa_y\,d\kappa_z\over(2\pi)^2}
{\lambda_n\over \nu^3}\int\limits_0^{\infty}
{t^3 e^{-t}\,dt\over 1 - \lambda_n e^{-t}}
={3\hbar c\over 8\pi d^4}\sum_{n=1}^2\int{d\kappa_y\,d\kappa_z\over(2\pi)^2}
{\mbox{Li}_4(\lambda_n)\over \nu^3},
\end{equation}
where the integral over $t$ results in the polylogarithm
$\mbox{Li}_4(z) = \sum_{n=1}^\infty z^n/n^4$.

Thus, the Casimir force in layers of moving (nondispersive) media has
the same dependence on the distance as the Casimir force between two
ideally conducting plates in vacuum. It is also seen that the value
and the sign of the force \r{force2} are determined by the the
eigenvalues $\lambda_{1,2}$ of the matrix $\=R_1\.\=R_2$, which in
turn depend on the relative velocities of the layers. In the next
section we will study numerically this dependence and will demonstrate
that the Casimir forces in moving media may be repulsive under certain
conditions.

\section{\label{numerics}Numerical examples and discussion}

In this section the expression for the Casimir force \r{force2} is
analyzed numerically. It is convenient to start from discussing some
properties of the reflection coefficients \r{reflmatrSec}. First of
all, we would like to remind that the elements of the reflection
matrix \r{reflmatrSec} are defined in terms of just a single component
of the electric and magnetic field vectors (see
Appendix~\ref{AppA}). Therefore, in general, their values differ
significantly from the classic reflection coefficients into co- and
cross-polarized TM and TE waves (the cases when \r{reflmatrSec}
reduces to the classic formulas are mentioned in
Appendix~\ref{AppA}). For example, the magnitudes of the
cross-components $R^{\rm eh}$ and $R^{\rm he}$ in our definition may
exceed unity when the characteristic impedance of the layers is
different from the free-space impedance $\eta_0$.

\begin{figure}
\centering
\epsfig{file=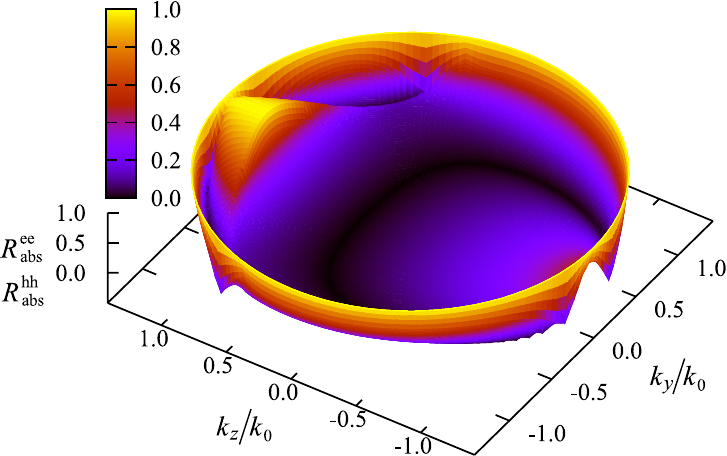,width=0.45\textwidth}
\caption{\label{figRee} (Color online) The absolute values of the reflection
  coefficients $R_{\rm abs}^{\rm ee} = |R^{\rm ee}|$ and $R_{\rm
    abs}^{\rm hh} = |R^{\rm hh}|$ (the respective plots coincide and
  are shown with a single surface) at the real frequencies as
  functions of the normalized wavenumbers $k_z/k_0$ and $k_y/k_0$ at
  an interface of a stationary medium with $\E_{\rm r}=2$, $\M_{\rm
    r}=1$ and the same medium moving with velocity $v = 0.6 c$. The
  plotted surface is colored proportionally to the reflection
  amplitude, as indicated in the color bar on the left.}
\end{figure}

\begin{figure}
\centering
\epsfig{file=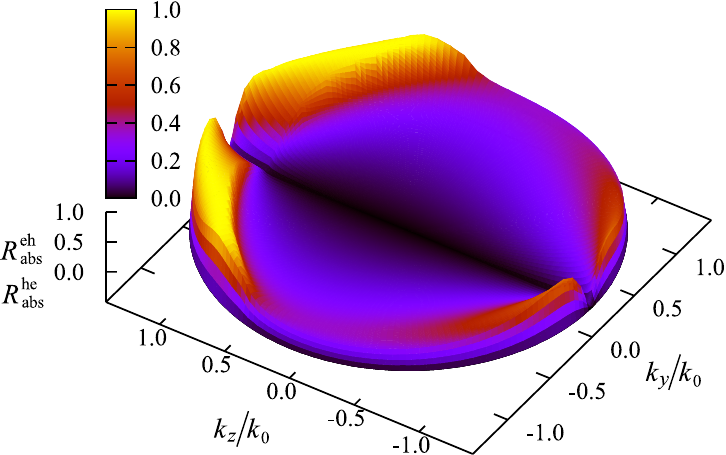,width=0.45\textwidth}
\caption{\label{figReh} (Color online) The absolute values of the normalized
  reflection coefficients $R_{\rm abs}^{\rm eh} = |R^{\rm
    eh}|\sqrt{\E_{\rm r}/\M_{\rm r}}$ and $R_{\rm abs}^{\rm he} =
  |R^{\rm he}|\sqrt{\M_{\rm r}/\E_{\rm r}}$ (the plots of these
  functions coincide). The parameters and the rest of the legend are
  the same as in Fig.~\ref{figRee}.}
\end{figure}

At real frequencies the elements of the reflection matrix~\r{reflmatrSec}
behave as shown in Figs.~\ref{figRee}--\ref{figReh}. In these figures
we plot the absolute values of the reflection coefficients at an
interface of a stationary medium with the relative parameters $\E_{\rm
  r} = 2$, $\M_{\rm r} = 1$ and the same medium moving with velocity
$v = 0.6c$ along the $z$-axis as functions of the relative wavenumbers
$k_y/k_0$ and $k_z/k_0$ (where $k_0=\o/c$) of an incident wave (the wave is incident from
the side of the stationary layer). The cross-components of the
reflection matrix plotted in Fig.~\ref{figReh} are normalized as
indicated in the figure caption. In these figures only the propagating
waves are considered, i.e., the waves with $(k_y/k_0)^2+(k_z/k_0)^2
\le \E_{\rm r}\M_{\rm r}$.

As one may notice, the elements of the reflection matrix demonstrate a
strongly nonreciprocal behavior: the reflection is different for the
incident waves with positive and negative $k_z$. It is also noticeable
that the reflection is rather low overall because the parameters of
the layers are chosen so that there would be no reflection if there
were no movement. However, the grazing waves reflect strongly, as well as
the waves with $k_z/k_0 \ge (\sqrt{\E_{\rm t}\M_{\rm t}} -
a)/\sqrt{\E_0\M_0} \approx 1.09$. The latter is due to the fact that
the waves with $k_z$ greater than the mentioned limit are evanescent
in the moving layer, as can be easily seen from the dispersion
equation.

\begin{figure}
\centering
\epsfig{file=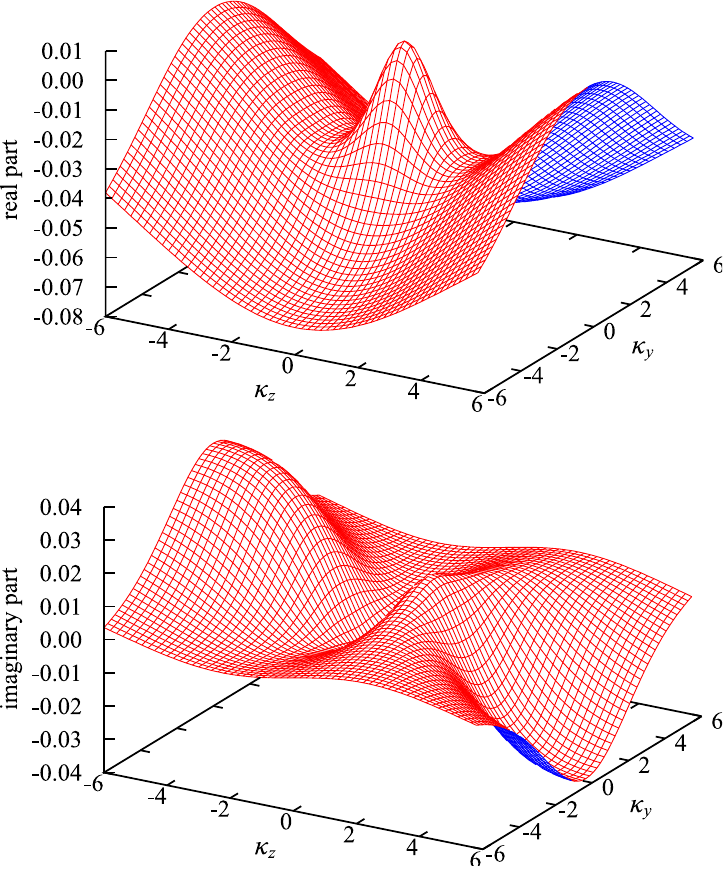,width=0.45\textwidth}
\caption{\label{figLambdaRep} (Color online) Real and imaginary parts of the
  eigenvalues $\lambda_{1,2}$ of the matrix
  $\overline{\overline{R}}_1\.\overline{\overline{R}}_2$ at imaginary
  frequencies as functions of the normalized wavenumbers $\kappa_z$
  and $\kappa_y$ (the plots for $\lambda_1$ and $\lambda_2$ coincide
  and are shown with a single surface). There are three layers of the
  same medium with $\E_{\rm r} = 2$, $\M_{\rm r} = 1$. The middle
  layer is stationary and the two outer layers move with the velocity
  $v = 0.6c$ along the positive direction of the $z$-axis.}
\end{figure}

\begin{figure}
\centering
\epsfig{file=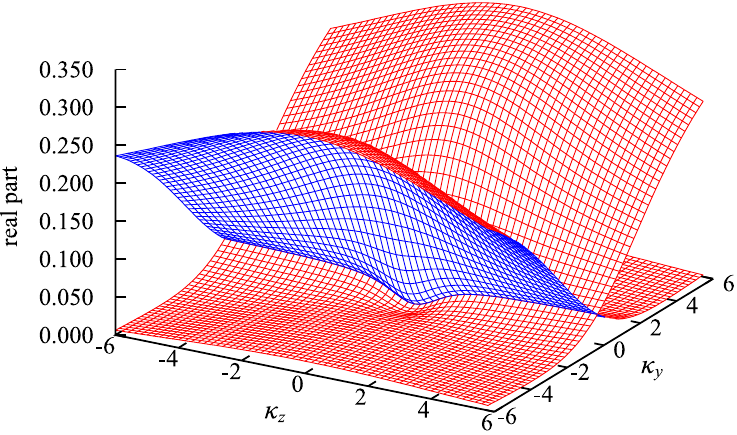,width=0.45\textwidth}
\caption{\label{figLambdaAtt} (Color online) The eigenvalues $\lambda_{1,2}$ at
  imaginary frequencies as functions of the normalized wavenumbers
  $\kappa_z$ and $\kappa_y$ for the case when the outer layers move in
  opposite directions (the eigenvalues are purely real in this
  scenario). The absolute value of the velocity and the other
  parameters are the same as in Fig.~\ref{figLambdaRep}.}
\end{figure}

The behavior at the imaginary frequencies is better understood from the
eigenvalues $\lambda_{1,2}$ of the matrix $\=R_1\.\=R_2$ written for
the complete structure composed of the three moving
layers. Accordingly to~\r{force2}, these eigenvalues determine the
magnitude and the sign of the Casimir force. The plots of the
eigenvalues are given in Fig.~\ref{figLambdaRep} for the case when the
outer layers move in the same direction with velocity $v = 0.6c$, and
in Fig.~\ref{figLambdaAtt} for the case when the two outer layers move
with the same speed, but in the opposite directions. The middle layer
is stationary in both cases.

In the case when the two outer layers move in the same direction with
the same velocity (Fig.~\ref{figLambdaRep}) the two eigenvalues of the
matrix $\=R_1\.\=R_2$ coincide. The eigenvalues are complex in this
case, with the real part concentrated mostly in the negative half
space, and the imaginary part changing sign when $k_z$ changes sign,
which is a consequence of the fact that
$\left(\=R_{1,2}(i\xi,k_y,k_z)\right)^* = \=R_{1,2}(-i\xi,k_y,k_z) =
\=R_{1,2}(i\xi,k_y,-k_z)$ when $\xi$, $k_y$ and $k_z$ are real.

\begin{figure}
\centering
\epsfig{file=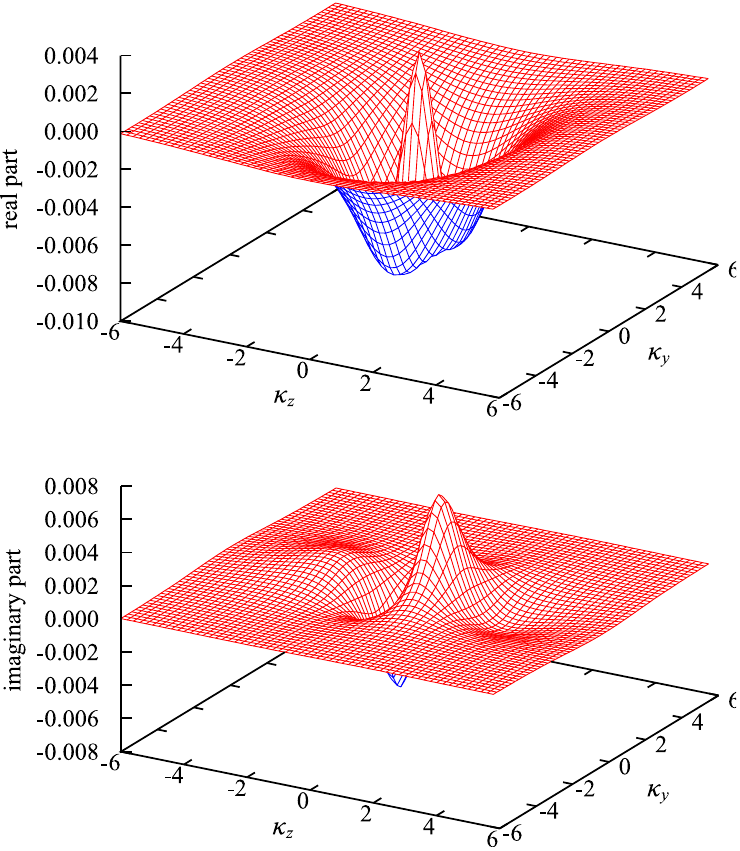,width=0.45\textwidth}
\caption{\label{figIntRep} (Color online) The real and the imaginary parts of the
  integrand of Eq.~\r{force2} as functions of the normalized
  wavenumbers $\kappa_z$ and $\kappa_y$ in the same scenario as in
  Fig.~\ref{figLambdaRep}.}
\end{figure}

\begin{figure}
\centering
\epsfig{file=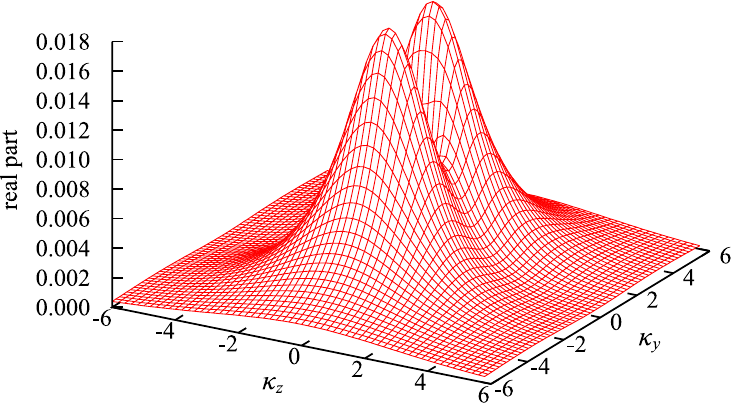,width=0.45\textwidth}
\caption{\label{figIntAtt} (Color online) The integrand of Eq.~\r{force2} as a function
  of the normalized wavenumbers $\kappa_z$ and $\kappa_y$ in the same
  scenario as in Fig.~\ref{figLambdaAtt} (the integrand is purely real
  in this scenario).}
\end{figure}

When substituted into the integral \r{force2} the dominating negative
real parts of the eigenvalues result in a negative Casimir force,
which corresponds to a {\em repulsion.} The contribution of the
imaginary part vanishes due the symmetry of the integrand. To further
illustrate this, in Fig.~\ref{figIntRep} we plot the integrand
of~\r{force2} as a function of the normalized wavenumbers $\kappa_z$
and $\kappa_y$. As is seen, only a small area of the
$(\kappa_z,\kappa_y)$ plane contributes to the integral, with the
negative values of the integrand on the periphery of this area clearly
outweighing the positive values seen at the middle.

When the two outer layers move in the opposite directions with the
same absolute speed (Fig.~\ref{figLambdaAtt}) the eigenvalues of the
matrix $\=R_1\.\=R_2$ are both real and positive (in a less symmetric
scenario when the absolute velocities of the two layers differ there
also appears a non-zero imaginary part). Thus, this case results in
{\em attraction} between the two moving layers, as clearly seen from
the plot of the integrand of~\r{force2} in Fig.~\ref{figIntAtt}. This
agrees with findings of Ref.~\cite{Leonhardt_nofriction_NJP_2009},
where only this type of relative movement of dielectric slabs
(separated by vacuum) was considered.

\begin{figure}
\centering
\epsfig{file=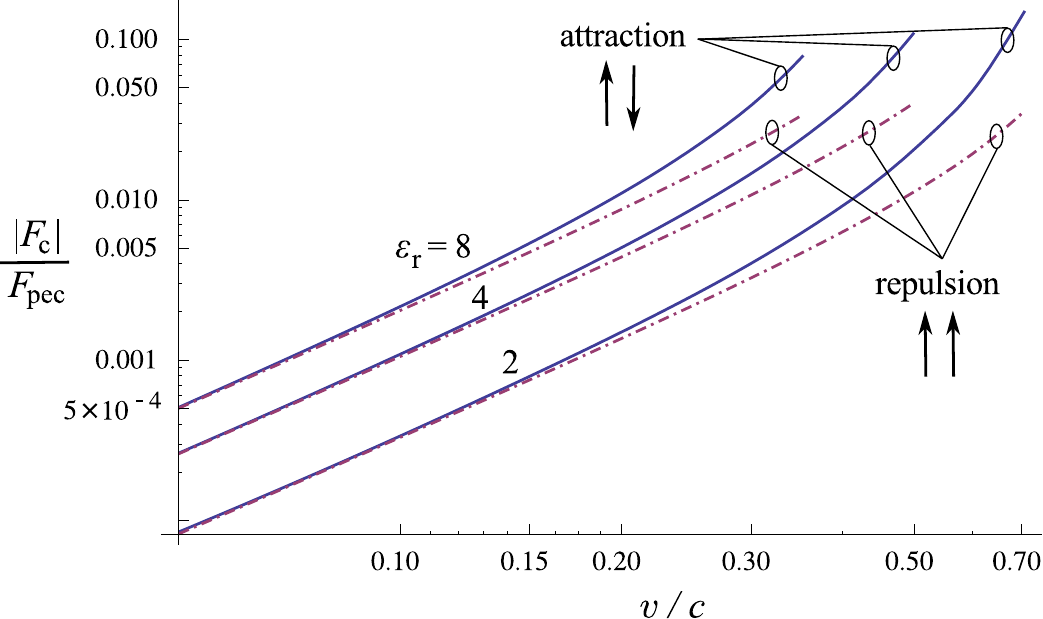,width=0.45\textwidth}
\caption{\label{figForce} (Color online) The magnitude of the attractive and
  repulsive Casimir-Lifshitz forces in the triple-layered structures
  with $\M_{\rm r} = 1$ and $\E_{\rm r}$ indicated in the plot, as
  functions of the relative velocity $v/c$ of the outer layers
  (logarithmic scale). The force is normalized to the Casimir force
  between two perfect electric conductors (PEC) separated by the same
  distance as the thickness of the middle layer. The arrows in the
  plot indicate the directions of the movement of the outer layers
  that result in attraction and in repulsion.}
\end{figure}

To further study the attraction and repulsion phenomena in moving
layers we have calculated the velocity dependence of the attractive
and repulsive Casimir-Lifshitz forces in the two scenario considered
above. The results are represented in Fig.~\ref{figForce}. In this
figure we plot the magnitude of the force $|F_{\rm c}|$ normalized to
the attractive Casimir force in a system of two ideally conducting
plates $F_{\rm pec} = \pi^2\hbar c/(240 d^4)$, where $d$ equals the
thickness of the middle (stationary) layer (as we noticed in
Section~\ref{layered} the dependence of the force on distance in
layers of moving nondispersive dielectrics is the same as in Casimir's
canonical structure). Fig.~\ref{figForce} also demonstrates the
dependence of the force on the value of the dielectric constant.

One can see that at low velocities the repulsive and attractive forces
in the two scenaria of the relative movement of the outer layers are
close to each other, while at larger speeds the attraction is
stronger than the repulsion. The double logarithmic scale of
Fig.~\ref{figForce} indicates that at small velocities both forces are
proportional to $(v/c)^2$, thus, the effect reported in this paper has
the same order as most of the relativistic effects. Quite naturally,
the effects are more pronounced in media with higher permittivity.

\begin{figure}
  \centering
  \epsfig{file=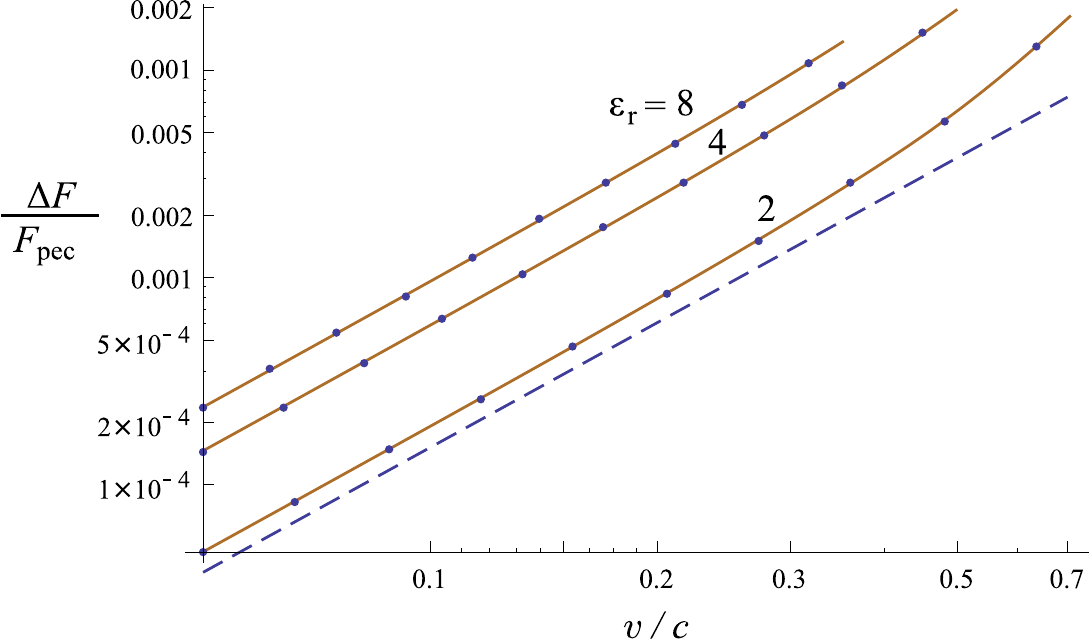,width=0.45\textwidth}
  \caption{\label{figleonhardt} (Color online) The additional attractive force $\Delta
    F = F(v) - F(v = 0)$ exerted on a dielectric with the relative
    permittivity $\E_{\rm r}$ moving with the relative velocity $v/c$
    nearby a stationary dielectric of the same permittivity, for the
    three different values of the relative permittivity: 2, 4,
    and~8. The dielectrics are separated by a vacuum gap. The force is
    normalized to the Casimir force between two stationary PEC plates
    separated by the same gap. The brown solid lines: the force
    calculated with the theory of the present paper
    [Eq.~\r{force2}]. The blue dots: the same force calculated from
    Eq.~(40) of Ref.~\cite{Leonhardt_nofriction_NJP_2009}. The blue
    dashed line: the plot of the $\beta^2$-proportional term of Eq.~(42)
    of Ref.~\cite{Leonhardt_nofriction_NJP_2009} for $\E_{\rm r} =
    2$.}
\end{figure}

In the last numerical example we calculate the attractive force
between a stationary and a moving dielectric separated by a vacuum
and compare it with the same force derived in
Ref.~\cite{Leonhardt_nofriction_NJP_2009} with an independent Green
tensor-based approach. The results of this comparison can be seen in
Fig.~\ref{figleonhardt}, where we plot $\Delta F = F(v) - F(v = 0)$
which is an addition to the force that appears because of the relative
movement of the layers.

From these calculations we conclude that up to the accuracy of
numerical integration (which is a triple integration in the case of
Ref.~\cite{Leonhardt_nofriction_NJP_2009}) the results of the two
independent approaches expressed by Eq.~\r{force2} of the present
paper and Eq.~(40) of Ref.~\cite{Leonhardt_nofriction_NJP_2009} are in
excellent agreement. The same reference contains also an expression
for the leading $O(\beta^2)$ term of the velocity-dependent correction
to the Lifshitz force (Eq.~(42) of
Ref.~\cite{Leonhardt_nofriction_NJP_2009}). However, one must be
accurate when making a comparison against this result, because the
(velocity-dependent) addends $A_{EE}^{-1}$ and $A_{BB}^{-1}$ seem to
appear there not expanded in powers of $\beta$. A plot of the explicit
$\beta^2$-proportional term~\footnote{The remaining velocity-dependent
  quantities in this term have been calculated at $\beta = 0$.} of the
mentioned expression is shown in Fig.~\ref{figleonhardt} with a blue
dashed line which does not match the exact result at low velocities.

Although it is out of the scope of this paper, the observed agreement
suggests that calculations of
Ref.~\cite{Leonhardt_nofriction_NJP_2009} are applicable to the
geometries that can be considered as effectively closed ones (which is
also the case of this paper) in which the pertinent difficulty with
the branch cuts pointed out in
Refs.~\cite{Pendry_friction_NJP_2010,Pendry_reply_NJP_2010} can be
treated in a manner similar to what we have done in
Appendix~\ref{AppB}.  Indeed, in this work we have shown that the
branch points of the reflection coefficients of moving layers are
irrelevant in such geometries.

\section{Conclusions}

In this paper we have considered the forces due to quantum-mechanical
fluctuations of the electromagnetic field in layered moving media. We
have demonstrated that rapid relative movements of neighboring layers
in a dielectric (e.g., in a nonuniform fluid flow) may result in both
attractive and repulsive interactions between the layers.

Although in the present study we have made an emphasis on the
Casimir-Lifshitz forces resulting from relativistic movement of
material layers, the results of this paper apply also (at least,
qualitatively) to a class of bianisotropic metamaterials called {\em
  moving media}. Thus, we may conclude that a specific type of {\em
  nonreciprocal} magnetoelectric interaction in bianisotropic
composites may also result in repulsive Casimir-Lifshitz
interactions. There have been previous attempts to realize Casimir
repulsion in metamaterials with the help of {\em reciprocal}
magnetoelectric interaction (e.g., chirality). However, it was
recently shown \cite{Silveirinha_norepulsion_PRB_2010,Silveirinha_restrictions_PRA_2010,
  Silveirinha_chiralcomment_PRL_2010} that the causality and passivity preclude Casimir
repulsion in reciprocal metamaterials.

The Casimir-Lifshitz interactions studied in this paper may be of
importance in areas of physics involving rapid movements of matter, as
well as in the phenomenological quantum electrodynamics of
nonreciprocal materials.

\section*{Acknowledgement}
The author is indebted to M\'{a}rio G. Silveirinha for fruitful
discussions and various suggestions, especially on the treatment of
the branch points of the reflection coefficients of the moving layers.

\appendix

\section{\label{AppB}}

The problem of branch points in the context of Casimir's energy
calculation dates back to 70's of the last century. Some of the main
ideas of the approach that we are going to use in this Appendix have
been borrowed from Ref.~\cite{Schram_argprinciple_PL_1973}.

\begin{figure}
\centering
\epsfig{file=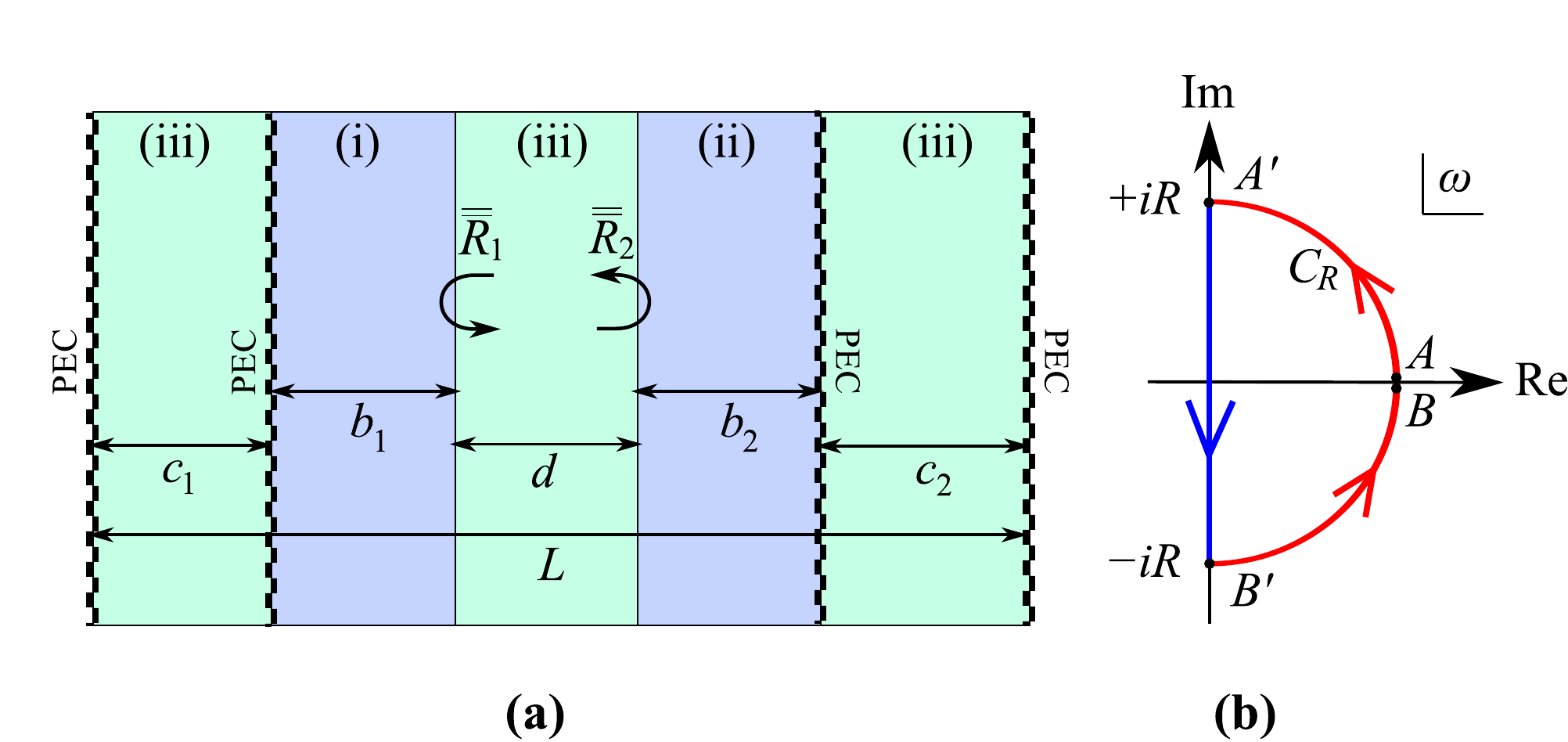,width=0.9\textwidth}
\caption{\label{appfig1}(Color online) (a) The PEC-backed structure used in calculation of the
  distant-dependent part of the zero-point energy. (b) The integration
  path $C$ in the complex plane of $\o$.}
\end{figure}

Instead of considering an initially open structure, we start with the
situation in which the moving layers are bounded by PEC walls, as
depicted in Fig.~\ref{appfig1}(a). As is seen, there are two PEC-backed
layers of media (i) and (ii) that can move in a background filled with
medium (iii) that is in turn terminated by two PEC walls at $x = 0$ and
$x = L$. We assume that $d \ll b_{1,2}$, and $b_{1,2} \ll L$. When the
PEC-backed layers (i) and (ii) move, their thicknesses $b_{1,2}$, as
well as the total size of the structure $L=c_1+b_1+d+b_2+c_2$, remain
fixed.

In this structure, the three regions $0<x<c_1$, $c_1<x<L-c_2$, and
$L-c_2<x<L$ are electromagnetically screened from each
other. Therefore, the characteristic equation for the whole structure
is a product of the equations for the three regions:
\begin{equation}
\l{appchareq}
  \tilde{{\cal D}}(\gamma) = \left(1-e^{-2\gamma c_1}\right)^2\x
\det\left\{\=I-\=R_1(\gamma)\.\=R_2(\gamma)e^{-2\gamma d}\right\}\x
\left(1-e^{-2\gamma c_2}\right)^2.
\end{equation}
In the middle of \r{appchareq} one can recognize the term that has the
form \r{chareq}; we have also made explicit the dependence of terms of
\r{appchareq} on the propagation factor in the background
medium  $\gamma$.

The reflection coefficients $\=R_{1,2}$ have the following
important property:
\begin{equation}
\l{rprop}
\=R_{1,2}(-\gamma)=\left[\=R_{1,2}(\gamma)\right]^{-1},
\end{equation}
which can be seen from the fact that the reflection dyadics can be
expressed as $\=R_{1,2}(\gamma)\equiv-\left[\=I+\=Z_{1,2}\.\=Y_{\rm
    w}(\gamma)\right]^{-1}\.\left[\=I-\=Z_{1,2}\.\=Y_{\rm
    w}(\gamma)\right] = -\left[\=I-\=Z_{1,2}\.\=Y_{\rm
    w}(\gamma)\right]\.\left[\=I+\=Z_{1,2}\.\=Y_{\rm
    w}(\gamma)\right]^{-1}$, where~$\=Z_{1,2}$ are the dyadic input
impedances of the PEC-backed layers which are meromorphic in the whole
complex plane of~$\o$ and independent of $\gamma$, and $\=Y_{\rm
  w}(\gamma)$ is the dyadic wave admittance of the middle layer that
is such that $\=Y_{\rm w}(-\gamma) = -\=Y_{\rm w}(\gamma)$. It should
be noted here that while the reflection matrix of an open half space
has the same property~\r{rprop}, the input impedance of such a space
is {\em not} a meromorphic function of
$\o$ (see Ref.~\cite{Schram_argprinciple_PL_1973}). Thus, we may conclude that
the branch points of $\=R_{1,2}$ coincide with the branch points of
$\=Y_{\rm w}$ that are at the frequencies where $\gamma(\o,k_y,k_z) =
0$.

Using the above property we may express $\tilde{{\cal D}}(-\gamma)$ in terms
of $\tilde{{\cal D}}(\gamma)$:
\begin{equation}
\l{dminusgam}
\tilde{{\cal D}}(-\gamma) = \left(1-e^{+2\gamma c_1}\right)^2
\det\left\{\=I-\=R_1^{\,-1}\.\=R_2^{\,-1}e^{+2\gamma d}\right\}
\left(1-e^{+2\gamma c_2}\right)^2 = 
{\tilde{{\cal D}}(\gamma)e^{4\gamma(L-b_1-b_2)}\over\det\left\{\=R_1\.\=R_2\right\}},
\end{equation}
where $\=R_{1,2}\equiv \=R_{1,2}(\gamma)$. From \r{dminusgam} one can
see that the roots of function $\tilde{\cal D}(-\gamma)$ in $\o$
include, in general, all the roots of $\tilde{\cal D}(\gamma)$. Thus,
we may construct a function
\begin{equation}
{\cal F}(\gamma) = \tilde{\cal D}(\gamma)\tilde{\cal D}(-\gamma) = 
{\tilde{\cal D}^2(\gamma)\,
e^{4\gamma(L-b_1-b_2)}\over\det\left\{\=R_1\.\=R_2\right\}} =
{{\cal D}^2(\gamma)\,
\left(1-e^{-2\gamma c_1}\right)^4\left(1-e^{-2\gamma c_2}\right)^4
e^{4\gamma(L-b_1-b_2)}
\over\det\left\{\=R_1\.\=R_2\right\}},
\end{equation}
where ${\cal D}(\gamma) \equiv
\det\left\{\=I-\=R_1(\gamma)\.\=R_2(\gamma)e^{-2\gamma d}\right\}$ has
the form \r{chareq}. The function ${\cal F}(\gamma)$ has all the roots
of ${\cal D}(\gamma)$ (with a difference that simple roots of ${\cal
  D}$ become roots of second order in ${\cal F}$) and is even in
$\gamma$. The latter makes ${\cal F}(\gamma)$ a meromorphic function
of~$\o$.

Therefore, we may apply the principle of argument (as explained in
Section~\ref{lifshitz}) to this function instead of applying it
directly to ${\cal D}(\gamma)$. The integral over the respective path (see Fig.~\ref{appfig1}(b)) in the
complex plane of $\o$ reads in this case
\begin{multline}
\l{appintF}
{1\over 4\pi i}\oint\limits_C\o\,d\log {\cal F} =
{1\over 2\pi i}\int\limits_{C_{AB}}\o\,d\log {\cal D} +
{1\over 4\pi i}\int\limits_{C_{AB}}\o\,d\log\left[
{\left(1-e^{-2\gamma c_1}\right)^4\left(1-e^{-2\gamma c_2}\right)^4}
\right] +\\
+{1\over 4\pi i}\int\limits_{C_{AB}}\o\,d\log
{e^{4\gamma(L-b_1-b_2)}\over\det\left\{\=R_1\.\=R_2\right\}},
\end{multline}
where $C_{AB}$ is an open path that is obtained from $C$ by
introducing a cut at the point where the semicircle crosses the real
axis. Such a cut is necessary because the expressions under the
integrals on the right hand side of \r{appintF} are not meromorphic in
$\o$.

Physically, the integral \r{appintF} represents a part of the zero
point energy that is due to the modal frequencies which are the roots
of \r{appchareq} that have been encircled by the path $C$. Because we
are interested only in the variation of the zero point energy with the
separation $d$ between the two moving slabs, we may drop the last
addend on the right hand side of~\r{appintF} as it is independent of
$d$. The second addend can be made arbitrary small when
$c_{1,2}\rightarrow\infty$ due to the nonvanishing positive real part
of $\gamma$. Thus, the distant-dependent part of the integral
\r{appintF} is given by
\begin{equation}
  \l{appintD}
  {1\over 2\pi i}\int\limits_{C_{AB}}\o\,d\log {\cal D} = -{1\over 2\pi i}\int\limits_{-iR}^{+iR}\o\,d\log {\cal D} +
  {1\over 2\pi i}\!\!\!\!\int\limits_{C_{AA'} \cup C_{B'B}}\!\!\!\!\o\,d\log {\cal D},
\end{equation}
where the first integral on the right hand side is taken over a path
that lies on the imaginary axis and the second integral is over the
two halves of the semicircle.

The integral \r{appintD} depends on the thicknesses of the
slabs~$b_{1,2}$ and the slab separation~$d$. Now we let $b_{1,2}
\rightarrow \infty$ in~\r{appintD} (when taking this limit, we assume
that still $L\gg b_{1,2}$). In this limit, due to the
nonvanishing imaginary part of $\o$ under the integrals on the right
hand side of \r{appintD}, the reflection coefficients of the
PEC-backed layers $\=R_{1,2}$ will tend to the respective reflection
coefficients of open half spaces (which are derived in
Appendix~\ref{AppA}).

The last step of the derivation is to let the radius of the semicircle
tend to infinity: $R\rightarrow\infty$. In this limit, which
corresponds to infinitely high frequencies, all dispersive materials
(including the materials with very weak dispersion that we consider in
this paper) become transparent. Therefore, $\=R_{1,2}\rightarrow 0$
under the integral over the semicircle, and this integral
vanishes. This leads to the expression~\r{lifshitz} for the
interaction part of the zero-point energy.

\section{\label{AppA}}

Let us consider an interface in a pair of layers. Without any loss of
generality we let the interface be at $x = 0$ with the $x$-axis
orthogonal to the interface. At the interface the tangential
components of the electric and magnetic fields of the two main
polarizations are given by Eqs.~\r{TMzH}--\r{TEzE}:
\begin{align}
\l{TMtan}
\begin{split}
H_y &= -{\o\E_{\rm t}k_x\over \o^2\E_{\rm t}\M_{\rm t}-(k_z+\o a)^2}E_z,\\
E_y &= -{k_y(k_z+\o a)\over \o^2\E_{\rm t}\M_{\rm t}-(k_z+\o a)^2}E_z,\\
H_z &= 0,
\end{split}
&(\mbox{TM}_z)
\end{align}

\begin{align}
\l{TEtan}
\begin{split}
E_y &= {\o\M_{\rm t}k_x\over \o^2\E_{\rm t}\M_{\rm t}-(k_z+\o a)^2}H_z,\\
H_y &= -{k_y(k_z+\o a)\over \o^2\E_{\rm t}\M_{\rm t}-(k_z+\o a)^2}H_z,\\
E_z &= 0,
\end{split}
&(\mbox{TE}_z)
\end{align}
where we have replaced $k_{\rm t}^2$ in the denominator with an equivalent
expression that follows from Eqs.~\r{TMz}--\r{TEz}. These
relations hold at both sides of the interface, with the material
parameters $\E_{\rm t}$, $\M_{\rm t}$, and $a$, and the wave vector
components taken at the respective sides.

In the following we are going to formulate and solve a plane wave
reflection problem at an interface of two moving media. To simplify
writing we introduce the following notations
\begin{align}
\l{alphabeta1}
\alpha  &= -{k_y(k_z+k_0a/\sqrt{\E_0\M_0})\over k_0^2\E_{\rm t}\M_{\rm t}/(\E_0\M_0)-(k_z+k_0a/\sqrt{\E_0\M_0})^2},\\
\beta^E &= {k_0k_x(\E_{\rm t}/\E_0)\over k_0^2\E_{\rm t}\M_{\rm t}/(\E_0\M_0)-(k_z+k_0a/\sqrt{\E_0\M_0})^2},\\
\beta^H &= {k_0k_x(\M_{\rm t}/\M_0)\over k_0^2\E_{\rm t}\M_{\rm t}/(\E_0\M_0)-(k_z+k_0a/\sqrt{\E_0\M_0})^2},
\l{alphabeta2}
\end{align}
where $k_0 = \o\sqrt{\E_0\M_0}$. Then, with these notations at hand we
consider a TM$_z$ wave of unit amplitude incident from the region $x <
0$ and write the fields in this region (the factor $e^{i(k_yy +
  k_zz)}$ common at both sides of the interface is dropped) as
\begin{align}
E_z &= e^{ik_x^{(1)}x} + A e^{-ik_x^{(1)}x},\\
H_z &= B e^{-ik_x^{(1)}x},\\
H_y &= -\eta_0^{-1}\beta_1^E e^{ik_x^{(1)}x} + \eta_0^{-1}\beta_1^E A e^{-ik_x^{(1)}x} + \alpha_1 B e^{-ik_x^{(1)}x},\\
E_y &= \alpha_1 e^{ik_x^{(1)}x} + \alpha_1 A e^{-ik_x^{(1)}x} - \eta_0\beta_1^H B e^{-ik_x^{(1)}x},
\end{align}
and in the region $x > 0$ as
\begin{align}
E_z &= C e^{ik_x^{(2)}x},\\
H_z &= D e^{ik_x^{(2)}x},\\
H_y &= -\eta_0^{-1}\beta_2^E C e^{ik_x^{(2)}x} + \alpha_2 D e^{ik_x^{(2)}x},\\
E_y &= \alpha_2 C e^{ik_x^{(2)}x} + \eta_0\beta_2^H D e^{ik_x^{(2)}x},
\end{align}
where $A$, $B$, $C$, and $D$ are yet unknown wave amplitudes of the
two reflected and the two transmitted waves, respectively, and
$\eta_0=\sqrt{\M_0/\E_0}$. As one can see, we take into account the
fact that a TM$_z$ incident wave may produce in general both
polarizations in the reflected and transmitted fields.

Equating the tangential components of the electric and magnetic fields
at both sides of the interface at $x\rightarrow 0$ one obtains a
system of four equations for the four unknown wave amplitudes. Solving
this system for $A$ and $B$ (i.e., for the reflected waves) we find
\begin{align}
\l{AandB}
A &= -{(\alpha_1-\alpha_2)^2-(\beta_1^E-\beta_2^E)(\beta_1^H+\beta_2^H)\over
(\alpha_1-\alpha_2)^2+(\beta_1^E+\beta_2^E)(\beta_1^H+\beta_2^H)},\\
B &= {2(\alpha_1-\alpha_2)\beta_1^E/\eta_0\over
(\alpha_1-\alpha_2)^2+(\beta_1^E+\beta_2^E)(\beta_1^H+\beta_2^H)}.
\end{align}

The case of a TE$_z$ incident wave can be considered in a completely
analogous manner. Below we give just the final result for the
amplitudes of the reflected waves:
\begin{align}
A' &= -{(\alpha_1-\alpha_2)^2-(\beta_1^H-\beta_2^H)(\beta_1^E+\beta_2^E)\over
(\alpha_1-\alpha_2)^2+(\beta_1^H+\beta_2^H)(\beta_1^E+\beta_2^E)},\\
B' &= -{2\eta_0(\alpha_1-\alpha_2)\beta_1^H\over
(\alpha_1-\alpha_2)^2+(\beta_1^H+\beta_2^H)(\beta_1^E+\beta_2^E)}.
\l{AandBprime}
\end{align}

Thus, we may introduce the following reflection matrix written in
terms of the $z$-components of the fields: \e \l{reflmatr} \vect{
  E_z^{\rm ref}\\
  \eta_0 H_z^{\rm ref} }=\matr{
  A & {\eta_0^{-1}B'}\\
  \eta_0 B & A' }\.  \vect{
  E_z^{\rm inc}\\
  \eta_0 H_z^{\rm inc} }\equiv \matr{
  R^{\rm ee} & R^{\rm eh}\\
  R^{\rm he} & R^{\rm hh} }\.  \vect{
  E_z^{\rm inc}\\
  \eta_0 H_z^{\rm inc} }.  \f As can be verified, the elements of the
reflection matrix reduce to the standard Fresnel reflection
coefficients of the P- and S-polarized waves in the special case of $a
= 0$, $k_y = 0$, for which $R^{\rm eh} = R^{\rm he} = 0$, $R^{\rm ee}
= R_{\rm p} \equiv
\left(\E_1k_x^{(2)}-\E_2k_x^{(1)}\right)/\left(\E_1k_x^{(2)}+\E_2k_x^{(1)}\right)$
and $R^{\rm hh} = -R_{\rm s} \equiv
\left(\M_1k_x^{(2)}-\M_2k_x^{(1)}\right)/\left(\M_1k_x^{(2)}+\M_2k_x^{(1)}\right)$,
and also in the case of $a = 0$, $k_z = 0$, for which $R^{\rm eh} =
R^{\rm he} = 0$, $R^{\rm ee} = R_{\rm s}$, and $R^{\rm hh} = -R_{\rm
  p}$. In the general case, the standard reflection matrix defined in
terms of the tangential components of the electric field can be
obtained from the matrix \r{reflmatr} with the following similarity
transformation:
\begin{equation}
\l{reflstd}
\matr{
  R^{yy} & R^{yz}\\
  R^{zy} & R^{zz}}
= 
\matr{\alpha_1 & -\beta^{H}_1\\
      1        & 0}\.
\matr{
  R^{\rm ee} & R^{\rm eh}\\
  R^{\rm he} & R^{\rm hh} }\.
\matr{\alpha_1 & -\beta^{H}_1\\
      1        & 0}^{-1}.
\end{equation}

As mentioned in Section~\ref{layered}, the reflection
coefficients \r{reflmatr} are in general complex, even at purely
imaginary frequencies. The complexity of the reflection matrix
\r{reflmatr} at imaginary frequencies is an unusual property that by
itself deserves a separate study. Here we will only briefly outline
the main reason behind this complexity. Indeed, from a physical point
of view, the reflection at imaginary frequencies $\o = i\xi$ can be
understood as the response to an incident wave that has the time
dependence of the form $e^{\xi t}$, i.e., to a signal that grows
exponentially with time. Let us now consider an interface between a
vacuum at $x < 0$ and a moving medium at $x > 0$, and let us assume
that there is a plane wave with such time dependence impinging on the
interface from the side of the the vacuum. We set up the same
coordinate system as above so that the movement is along the
$z$-axis. In this coordinate system the incident wave of, for
instance, the TM$_z$ polarization can be written as \e \l{illustr}
E_z^{\rm inc} = E_0 e^{i(k_y y + k_z z)}e^{\xi t - \gamma x} , \f
where $k_y$ and $k_z$ are the {\em real} propagation factors in the
interface plane, and $\gamma = -ik_x =
\sqrt{\xi^2\E_0\M_0+k_y^2+k_z^2} \ge 0$ is the solution of the vacuum
dispersion equation at imaginary frequencies. As we are interested
only in an illustration, we let $k_y = 0$ in~\r{illustr}, so that the
TM$_z$ wave becomes the standard TM wave with respect to the plane of
incidence.

A vacuum is invariant under the Lorentz transformations (see
Section~\ref{waves}), as are the components of the electromagnetic
fields parallel to the velocity vector (the $z$-components), therefore
to solve for the reflection coefficient we may switch to the comoving
frame in which the reflection coefficient is simply \e \l{illustr2}
R^{\rm ee} =
{\E_0\sqrt{(\xi')^2\E\M+(k_z')^2}-\E\sqrt{(\xi')^2\E_0\M_0 + (k_z')^2}
  \over \E_0\sqrt{(\xi')^2\E\M+(k_z')^2}+\E\sqrt{(\xi')^2\E_0\M_0 +
    (k_z')^2}}, \f where \e \l{illustr3}\xi'={\xi + i k_z v \over
  \sqrt{1-v^2/c^2}}, \quad k_z'={k_z - i \xi v/c^2 \over
  \sqrt{1-v^2/c^2}} \f are the imaginary frequency and the
$z$-component of the wave vector transformed to the comoving
frame. Substituting~\r{illustr3} into~\r{illustr2} we obtain after
some manipulation \e\l{illustr4} R^{\rm ee} =
{\sqrt{(n^2-1)(\xi'/c)^2+\gamma^2}-\E_{\rm r}\gamma \over
  \sqrt{(n^2-1)(\xi'/c)^2+\gamma^2}+\E_{\rm r}\gamma}, \f where
$\gamma = \sqrt{\xi^2\E_0\M_0 + k_z^2}$ and $\E_{\rm r} = \E/\E_0$. As
is readily seen, $R^{\rm ee}$ is in general complex when $n^2 \neq 1$
and $v \neq 0$, and this complexity is due to the fact that the
relative movement of the layers intermixes the imaginary frequencies
with the real-valued wavenumbers by the virtue of the Lorentz
transformations. It is easy to check that the result \r{illustr4} is a
particular case of more general formulas \r{AandB}--\r{reflmatr}.

Conversely, one may verify that if there exists a reference frame at
which the moving matter is at rest, then under a transformation of the
form~\r{illustr3} the complex propagation factor
$\gamma(i\xi,k_y,k_z)$ and the reflection matrix~\r{reflmatr} reduce
to the respective expressions in stationary
magnetodielectrics. Additionally, when such a transformation is
applied to the integrand of~\r{lifshitz}, one may notice that the
integration element $dk_y dk_z d\xi$ is preserved, because the
Jacobian of the transformation~\r{illustr3} equals unity:
$\partial(\xi,k_z)/\partial(\xi',k_z') = 1$. Therefore, the Casimir
force per unity of area (the Casimir pressure) given by~\r{lifshitz}
is the same in all reference frames that move parallel to the layers,
provided that the velocities of the layers are transformed accordingly
to the relativistic velocity addition law. Such an invariance of the
Casimir pressure~\r{lifshitz} is not surprising, as physically the
pressure exerted on the moving layers is related with the component of
the photon momenta that is perpendicular to the direction of the
movement, and this component is preserved under the Lorentz
transformation. Thus, we may conclude that our formulation extends the
known theory of Casimir-Lifshitz forces in dielectric layers in a way
fully consistent with special relativity.

\section{\label{AppC}}

In this appendix we discuss how the results obtained for
non-dispersive moving media may be generalized to include the effects
of frequency dispersion in the effective material parameters.

Let us consider an isotropic dispersive magnetodielectric described by
the following material relations in its proper frame: \e
\l{mr1}\_D'(\_x',t') = \E_0\int\limits_0^\infty{\E_{\rm
    r}(\tau')\_E'(\_x',t'-\tau')}\,d\tau', \f \e \l{mr2} \_B'(\_x',t')
= \M_0\int\limits_0^\infty{\mu_{\rm
    r}(\tau')\_H'(\_x',t'-\tau')}\,d\tau', \f where $\E_{\rm
  r}(\tau')$ and $\mu_{\rm r}(\tau')$ are the dielectric and magnetic
response functions.

In the proper frame which is co-moving with the medium, the field
components orthogonal to $\_v$ can be expressed through the same
components in the stationary laboratory frame as \e \l{applor1}\_E_{\rm
  t}' = \gamma_{\rm L}(\_E_{\rm t} + \_v\x\_B_{\rm t}), \quad \_H_{\rm t}' =
\gamma_{\rm L}(\_H_{\rm t} - \_v\x\_D_{\rm t}), \f \e \l{applor2}\_D_{\rm t}' =
\gamma_{\rm L}(\_D_{\rm t} + {1\over c^2}\_v\x\_H_{\rm t}), \quad \_B_{\rm t}'
= \gamma_{\rm L}(\_B_{\rm t} - {1\over c^2}\_v\x\_E_{\rm t}), \f where $\_v$
is the medium velocity (along $Oz$) and $\gamma_{\rm L} =
1/\sqrt{1-v^2/c^2}$. Substituting \r{applor1}--\r{applor2} into
\r{mr1}--\r{mr2} one obtains \e \_D_{\rm t} + {1\over c^2}
\_v\x\_H_{\rm t} = \E_0\int\limits_0^\infty{\E_{\rm
    r}(\tau')\left[\_E_{\rm t}(z(z',t'),t(z',t'-\tau')) +
    \_v\x\_B_{\rm t}(z(z',t'),t(z',t'-\tau')) \right] }\,d\tau', \f \e
\_B_{\rm t} - {1\over c^2} \_v\x\_E_{\rm t} =
\M_0\int\limits_0^\infty{\M_{\rm r}(\tau')\left[\_H_{\rm
      t}(z(z',t'),t(z',t'-\tau')) - \_v\x\_D_{\rm
      t}(z(z',t'),t(z',t'-\tau')) \right] }\,d\tau', \f where $z =
z(z',t') = \gamma_{\rm L}(z' + vt')$, $t = t(z',t') = \gamma_{\rm L}(t' +
vz'/c^2)$. From here, \e \l{int1} \_D_{\rm t} + {1\over c^2}
\_v\x\_H_{\rm t} = \E_0\int\limits_0^\infty{\E_{\rm
    r}(\tau')\left[\_E_{\rm t}(z - \gamma_{\rm L} v\tau',t - \gamma_{\rm L}\tau') +
    \_v\x\_B_{\rm t}(z - \gamma_{\rm L} v\tau',t - \gamma_{\rm L}\tau') \right]
}\,d\tau', \f \e \l{int2} \_B_{\rm t} - {1\over c^2} \_v\x\_E_{\rm t}
= \M_0\int\limits_0^\infty{\M_{\rm r}(\tau')\left[\_H_{\rm t}(z -
    \gamma_{\rm L} v\tau',t - \gamma_{\rm L}\tau') - \_v\x\_D_{\rm t}(z - \gamma_{\rm L}
    v\tau',t - \gamma_{\rm L}\tau') \right] }\,d\tau'.  \f

In order to obtain the constitutive relations in the laboratory frame,
one has to solve the system of integral equations
\r{int1}--\r{int2} to express $\_D$ and $\_B$ in terms of $\_E$ and
$\_H$. It is evident that, in general, the above system may not result
in a simple proportionality relation between the flux and field
vectors. However, for plane waves this system is easily solvable and
results in relations \r{matrel1}--\r{lor2} of Section~II with $\E =
\E(\o')$, $\M = \M(\o')$, and $n = n(\o')$, where $\o' = \gamma_{\rm
  L}(\o - k_zv)$ is the angular frequency in the proper frame of the
moving medium. As this frequency depends on the wavenumber in the
laboratory frame, the relations \r{matrel1}--\r{lor2} with the
modified parameters readily describe a spatially nonlocal medium, as
was mentioned in Introduction.

One may also verify that such modification does not affect the
frequency domain treatment of Section II. The equations
\r{TMz}--\r{TEzDB} written for the plane waves in a moving
nondispersive magnetodielectric hold also in the case of dispersive
moving media if the parameters $\E_{\rm t} = \E_{\rm t}(\o')$,
$\M_{\rm t} = \M_{\rm t}(\o')$, and $a = a(\o')$ are understood as
$\o'$-dependent. Eq.~\r{roots} becomes a transcendental equation in
the dispersive case. The important symmetry of Eqs.~\r{TMz}--\r{TEzDB}
with respect to the simultaneous change of signs of $\o$ and $\_k$
discussed in Section II is preserved in the dispersive case, because
$\o'(\o, \_k) = -\o'(-\o,-\_k)$ and $\E(-\o') = \E^*(\o')$, $\M(-\o')
= \M^*(\o')$. Thus, the generalization of the classical part of this
study to the dispersive case is trivial.

The quantum-theoretical part of this paper is based on the
expressions~~\r{hamfinal}--\r{hamsep} and \r{hamfinop} for the
Hamiltonian of the free electromagnetic field. As has been mentioned in
Section~\ref{hamiltonian}, these expressions are physically understood
as summations over the energies of all possible modes in a modal
expansion of the electromagnetic field. Therefore, it is only natural
that the same expressions must also hold in the case of frequency
dispersive material parameters, provided that the basic relations for
the energy $w$ and the momentum $p$ of a photon in a dispersive
medium remain the same as in a vacuum: $w = \hbar\o$, $p = \hbar k$,
$w/p = \o/k = v_{\rm ph}$. Hence, one must also expect the
diagonalized form of the Hamiltonian~\r{hama} to be valid in the
dispersive case, in which the modal frequencies $\o(\_k)$ are found
from the transcendental equation~\r{roots} that must take into account
the material dispersion.

Therefore, the expressions for the interaction part of the zero-point
energy~\r{lifshitz} and the Casimir force~\r{force1}--\r{force11} must also hold
in the dispersive case.


\let\o\oslashed


\end{document}